\pdfoutput=1
\RequirePackage{ifpdf}
\ifpdf
\documentclass[pdftex]{sigma}
\else
\documentclass{sigma}
\fi

\numberwithin{equation}{section}

\newcommand{\Hf}{\mathop\mathrm{Hf}\nolimits}
\newcommand{\Pf}{\mathop\mathrm{Pf}\nolimits}
\newcommand{\Res}{\mathop\mathrm{Res}\nolimits}
\newcommand{\Ber}{\mathop\mathrm{Ber}\nolimits}
\newcommand{\Z}{\mathbb{Z}}
\newcommand{\INZ}[2]{\in\frac#1#2 +\mathbb{Z}}
\newcommand{\C}{\textsf{C}_}

\newcommand{\G}{\Gamma}
\newcommand{\bt}{{\bf {t}}}

\renewcommand{\l}{\langle}
\renewcommand{\r}{\rangle}

\begin{document}

\allowdisplaybreaks

\renewcommand{\thefootnote}{$\star$}

\renewcommand{\PaperNumber}{036}

\FirstPageHeading

\ShortArticleName{CKP Hierarchy, Bosonic Tau Function and Bosonization Formulae}

\ArticleName{CKP Hierarchy, Bosonic Tau Function\\ and Bosonization Formulae\footnote{This
paper is a contribution to the Special Issue ``Geometrical Methods in Mathematical Physics''. The full collection is available at \href{http://www.emis.de/journals/SIGMA/GMMP2012.html}{http://www.emis.de/journals/SIGMA/GMMP2012.html}}}

\Author{Johan W.~VAN DE LEUR~$^\dag$, Alexander Yu. ORLOV~$^{\ddag}$ and Takahiro SHIOTA~$^\S$}
\AuthorNameForHeading{J.W.~van de Leur, A.Yu.~Orlov and T.~Shiota}
\Address{$^\dag$~Mathematical Institute, University of Utrecht,\\
\hphantom{$^{\dag}$}~P.O.~Box 80010, 3508 TA Utrecht, The Netherlands}
\EmailD{\href{mailto:J.W.vandeLeur@uu.nl}{J.W.vandeLeur@uu.nl}}
\Address{$^{\ddag}$~Nonlinear Wave Processes Laboratory, Oceanology Institute,\\
\hphantom{$^{\ddag}$}~36 Nakhimovskii Prospect, Moscow 117851, Russia}
\EmailD{\href{mailto:orlovs55@mail.ru}{orlovs55@mail.ru}}
\Address{$^\S$~Mathematics Department, Faculty of Science, Kyoto University, Kyoto 606-8502, Japan}
\EmailD{\href{mailto:shiota@math.kyoto-u.ac.jp}{shiota@math.kyoto-u.ac.jp}}

\ArticleDates{Received January 17, 2012, in f\/inal form June 07, 2012; Published online June 22, 2012}

\Abstract{We develop the theory of CKP hierarchy introduced in the papers
of Kyoto school [Date~E., Jimbo~M., Kashiwara~M., Miwa~T., \textit{J.~Phys. Soc. Japan} \textbf{50} (1981), 3806--3812]
(see also [Kac V.G., van~de Leur J.W.,   \textit{Adv. Ser. Math. Phys.}, Vol.~7, World Sci. Publ., Teaneck, NJ, 1989,
  369--406]). We
present appropriate bosonization formulae. We show that in the
context of the CKP theory certain orthogonal polynomials appear.
These polynomials are polynomial both in even and odd (in
Grassmannian sense) variables.}


\Classification{17B65; 17B67; 17B69; 20G43; 81R12}

\vspace{-2mm}

\renewcommand{\thefootnote}{\arabic{footnote}}
\setcounter{footnote}{0}

\section{Introduction}

\looseness=-1
In this paper we develop the ideas of  Date, Jimbo, Kashiwara
and Miwa \cite{DJKM-CKP,DJKMVI}. In
\cite{DJKMVI} it was pointed out that the tau function of the
CKP hierarchy may be presented as the vacuum expectation value of
bosonic f\/ields $\phi_n$ which act in a bosonic Fock space, denoted
by $F$ in the present paper. As it was shown in \cite{DJKMVI}
the higher CKP f\/lows are induced by the action of bosonic current
algebra operators $J_n$, $n>0$. We shall show that in contrast to
the familiar fermionic approach the action of the currents
$J_{-n}$ on the vacuum state does not generate the whole
Fock space $F$ where the original bosonic operators $\phi_n$ act.
To generate the whole Fock space we need to add an additional
fermionic f\/ield which is a sort of super-counterpart to the
bosonic current. This problem is studied  in Section~\ref{CKPbosonic}. In Subsection \ref{Bose-Fermi} we introduce
Fermi operators  whose action on the vacuum vector
complete the action of the current algebra $J_n$ to obtain the
whole Fock space~$F$. A bosonization formula which expresses the
original bosonic f\/ield $\phi(z)$ in terms of the current algebra  $J_n$
and of the fermion f\/ield $\theta(z)$ is suggested (\ref{theta}).
(As a byproduct of this relation we obtain an equality~(\ref{pfaffhaf}) which relates Pfaf\/f\/ian and Hafnian expressions
which earlier appeared in a quite dif\/ferent context \cite{IKO}.)
Here we show that this fermionic f\/ield is a~superpartner of
currents and it naturally creates a dependence of CKP tau function
on auxiliary odd Grassmannian parameters. Though we present a~bilinear equations written in terms of super vertex operator we do
not construct Lax equations with respect to odd parameters. In
Section~\ref{Related} we introduce new orthogonal polynomials in
many variables, $\C\lambda(\bt)$, which appear as a~result of the
`bosonization' of the basis Fock vectors of $F$, see~(\ref{C-lambda}). These polynomials depend both on CKP higher
times $t_n$ ($n$ odd) and the above-mentioned Grassmannian odd
parameters. In certain sense these polynomials play a role similar
to the role of the Schur functions in the theory of KP~\cite{Sa}
and TL~\cite{Tinit}, and the role of the projective Schur
functions in the theory of BKP~\cite{Nimmo, You}, namely,
CKP tau function may be presented as a series in these polynomials
over partitions (see Subsection~\ref{CKPpolynomials}). However in
contrast to the KP and BKP cases these polyno\-mials are not CKP tau
functions themselves. At the end some combinatorial properties of~$\C\lambda$ are discussed.

\section{CKP bosonic tau function\label{CKPbosonic}}

In this section we follow a suggestion in~\cite{DJKMVI} and describe a CKP hierarchy of PDEs starting from  a collection of free bosons.
This imitates their approach in the BKP case, where one starts with neutral fermions.
However, this hierarchy, although related to Lie algebra $c_\infty$,
dif\/fers from the usual CKP hierarchy, for which one takes
a reduction of the KP hierarchy by assuming that the Lax operator
satisf\/ies $L^*=-L$; such Lax operators come from certain KP tau functions,
which are f\/ixed by some involution and where one puts the even times to zero,
see e.g., \cite{DJKMVI} or \cite{AL} for more details.
The hierarchy described in this paper is dif\/ferent and is not related to the
usual CKP which comes from a reduction of KP. In the latter case
one has a realization of $c_\infty$, for which the level is positive.
Our construction realizes  $c_\infty$ with a {\it negative\/} level.

\subsection[Bose-Fermi correspondence in the CKP case]{Bose--Fermi correspondence in the CKP case}\label{Bose-Fermi}

We follow a suggestion of Date, Jimbo, Kashiwara and Miwa in their paper \cite{DJKMVI}
and introduce free bosons, but for convenience of notation we
shift the index by $\frac 12$. So
$\phi_i$ with $i\in\frac 12+\mathbb{Z}$ satisfy commutation relations:
\begin{gather}
\label{S1}
\phi_{i}\phi_j-\phi_j\phi_i=(-)^{j-\frac{1}{2}}\delta_{i,-j}.
\end{gather}
The Fock space $F$, respectively $F^*$, is def\/ined
by
\begin{gather}
\label{S2} \phi_j\, |0\rangle=0\quad\mbox{if}\ \ j<0,\qquad\mbox{resp.}\qquad
\langle 0|\, \phi_j=0\quad\mbox{if}\ \ j>0,
\end{gather}
so that $F$ has as basis the vectors
 \begin{gather}\label{S2BASIS}
(\phi_{j_1})^{m_1}(\phi_{j_{2}})^{m_{2}}
\cdots(\phi_{j_{n-1}})^{m_{n-1}}(\phi_{j_n})^{m_n}|0\rangle
 \end{gather}
with $j_1>j_2>\cdots> j_{n-1}>j_n>0$ and $m_i$ positive integers.
Def\/ining
\[
\deg |0\rangle =0 ,\qquad\deg\phi_j=j ,
\]
we have a direct sum decomposition of $F$:
\[
F=\bigoplus_{k\in\frac12\mathbb Z} F_k\qquad\mbox{with}\quad
F_k=\{f\in F\mid \deg f=k\} .
\]
It is straightforward to check that the dimension of $F_k$ is
given by the partition of $k$ into
 positive elements of $\frac 12+\mathbb{Z}$.
Def\/ine the formal character as
\[
\dim_q F=\sum_{k\in\frac12\mathbb Z} \dim F_k\, q^k .
\]
Then
\begin{gather}\label{q1}
\dim_q F=\prod_{0<k\INZ12}\frac 1{1-q^k} .
\end{gather}
Writing
\[
\phi(z)=\sum_{j\INZ12}\phi_jz^{j-\frac 12} ,
\]
we denote
\begin{gather}
\label{S4} H(z):=\sum_{n\in
1+2\mathbb{Z}}J_nz^{-n-1}:=-\frac 12\,{:}\phi(-z)\phi(z){:} ,
\end{gather}
where the normal ordering is def\/ined by
\begin{gather}\label{normal ordering}
{:}\phi_i\phi_j{:}=
\begin{cases}
\phi_i\phi_j&\mbox{if }i\ge j ,\\[2pt]
\phi_j\phi_i&\mbox{if }j>i .
\end{cases}
\end{gather}
In other words  $J_n=0$ for $n$ even and
\[ % \begin{gather} \label{S}
J_n=\frac{1}{2}\sum_{j\in
\frac{1}{2}+\mathbb{Z}}(-)^{j+\frac{1}{2}}\phi_j\phi_{-j-n}\qquad
\mbox{for\,\ $n$\,\ odd;}
\] % \end{gather}
one has the following familiar commutation relations
\begin{gather}
\label{S5} [J_n, J_m]=-\frac{n}{2}\delta_{m,-n} .
\end{gather}
The elements ${:}\phi_i\phi_j{:}$ form a representation of the
Lie algebra $c_\infty$, see e.g.~\cite{Kac-Leur}. However,
we want to stress that its level (the value of its central element) is negative.
Note also that in the commutation relations~(\ref{S5}) we
have the factor $-\frac{n}{2}$ instead of the usual~$\frac{n}{2}$.

It is clear that
\[
J_n|0\rangle=\langle 0|J_{-n}=0\qquad\mbox{for}\ \ n>0 .
\]
By a similar argument as before,
again since these are bosons, we can apply an element $J_{-n}$
inf\/initely many times to $|0\rangle$. Since the degree of $J_n$ is $-n$
 we obtain that the action of this Heisenberg algebra on the vacuum
vector produces in the $\dim_q F$ the partition function of partitions in only odd numbers:
\[
\prod_{0< k\in 1+2\mathbb{Z}}\frac{1}{1-q^k}.
\]
Now we calculate, using \eqref{q1},
\begin{gather}
\left(\prod_{0< k\in
1+2\mathbb{Z}}\frac{1}{1-q^k}\right)^{-1}\dim_q F=
\prod_{0< k\in\frac{1}{2}+\mathbb{Z}}\frac{{1-q^{2k}}}{1-q^k}
=\prod_{0< k\in\frac{1}{2}+\mathbb{Z}}\big(1+q^k\big).\label{q3}
\end{gather}
 This part should be explained by something else and we expect it to be
fermions, at least anticommuting variables. The factor~$1+q^k$
is related to a fermion of degree~$k$. This is how we
get these elements and calculate their commutation relations.

We f\/irst calculate
\begin{gather}
  [J_n, \phi(z)] =\frac 12\sum_{j,k\INZ12}(-)^{j+\frac 12}
[\phi_j\phi_{-j-n}, \phi_k]z^{k-\frac 12}\nonumber\\
\hphantom{[J_n, \phi(z)]}{} =\frac{1}{2}\sum_{j,k\INZ12}(-)^{j+\frac 12} \left(
[\phi_j, \phi_k]\phi_{-j-n}+ \phi_j[\phi_{-j-n}, \phi_k]
\right) z^{k-\frac12}\nonumber\\
\hphantom{[J_n, \phi(z)]}{}
=\frac 12\sum_{j,k\INZ12}(-)^{j+\frac 12}
(-)^{k-\frac 12}\left(\delta_{j,-k}\phi_{-j-n}+\delta_{j+n,k}\phi_j\right)
z^{k-\frac 12}\nonumber\\
\hphantom{[J_n, \phi(z)]}{}
=\frac12\sum_{k\INZ12} 2\phi_{k-n}z^{k-\frac 12}
=z^n\phi(z) .\label{S6}
\end{gather}
Now, using (\ref{S5}) we see that
\begin{gather}\label{S5bis}
\big[J_n,e^{\frac2m J_mz^{-m}}\big]=\delta_{n,-m}z^n e^{\frac2m J_mz^{-m}}.
\end{gather}
Hence setting
\begin{gather}\label{theta}
 \theta(z):=V_-(z)^{-1}\phi(z)V_+(z)^{-1},
\end{gather}
where
\begin{gather}
\label{exp}
V_\pm(z)
=\exp\sum_{\pm k>0,\,\mathrm{odd}}\frac{2}{k}J_k z^{-k} ,
\end{gather}
we have, from
\eqref{S6} and \eqref{S5bis},
\begin{gather}\label{comm-Htheta}
[J_n, \theta(z)]=0 .
\end{gather}

\subsection[Commutation relations of the $\theta(z)$'s]{Commutation relations of the $\boldsymbol{\theta(z)}$'s}

We now want to calculate the commutation relations of these
$\theta(z)$'s given in (\ref{theta}).  For this we f\/irst rewrite the commutation relations
(\ref{S1}) as follows:
\[
\phi(z)\phi(y)-\phi(y)\phi(z)=\delta(z-(-y))\,,
\]
where
\[
\delta(z-y)=z^{-1}\sum_{k\in\mathbb{Z}}
\left(\frac{z}{y}\right)^k.
\]
Note also that
\[
\phi(-z)\phi(y)={:}\phi(-z)\phi(y){:}-\frac{1}{z}\frac{1}{1-\frac{y}{z}}  .
\]
We f\/irst show the following identities
\begin{gather}
V_+(-z)^{-1}V_-(y)^{-1} =
\frac{1-\frac{y}{z}}{1+\frac{y}{z}}V_-(y)^{-1}V_+(-z)^{-1},\nonumber\\
\phi(-z)V_-(y)^{-1} =
\frac{1+\frac{y}{z}}{1-\frac{y}{z}}V_-(y)^{-1}\phi(-z),\nonumber\\
V_+(-z)^{-1}\phi(y) =
\frac{1+\frac{y}{z}}{1-\frac{y}{z}}\phi(y)V_+(-z)^{-1}.\label{ident}
\end{gather}
Introduce
\[
V(t)=\exp\sum_{k>0,\,\mathrm{odd}}t_k J_k .
\]
Then
\begin{gather}
\label{VT1}
V_+(z)=V\left(\frac21 z^{-1},\frac23 z^{-3},\frac25 z^{-5},\ldots\right) .
\end{gather}
The f\/irst equation of (\ref{ident}) is obtained in the following
way. First using \eqref{S5} one has
\begin{gather*}
V(t)V_-(y) =\exp \left[  \sum_{k>0,\,\mathrm{odd}} t_kJ_k,
- \sum_{\ell>0,\,\mathrm{odd}} \frac{2}{\ell}y^{\ell}J_{-\ell}\right]V_-(y)V(t)\\
\hphantom{V(t)V_-(y)}{}  =\exp \left(\sum_{k>0,\,\mathrm{odd}}t_{k}y^k
\right)
V_-(y)V(t)  .
\end{gather*}
Combining this with (\ref{VT1}) one obtains
\begin{gather*}
V_+(-z)^{-1}V_-(y)^{-1} =\exp \left(-\sum_{k>0,\,\mathrm{odd}} \frac{2}{k}\left(
\frac{y}{z}\right)^k\right)
V_-(y)^{-1}V_+(-z)^{-1}\\
\hphantom{V_+(-z)^{-1}V_-(y)^{-1}}{}
 =\frac{1-\frac{y}{z}}{1+\frac{y}{z}}V_-(y)^{-1}V_+(-z)^{-1}.
\end{gather*}
Using $\phi(-z)J_n=(J_n-(-z)^n)\phi(-z)$, see~\eqref{S6},
we obtain the second relation in \eqref{ident} as follows:
\begin{gather*}
\phi(-z)V_-(y)^{-1} =
\phi(-z)\exp\left(\sum_{k>0,\,\rm odd}\frac2kJ_{-k}y^k\right)
 =
\exp\left(\sum_{k>0,\,\rm odd}\frac2k(J_{-k}+z^{-k})y^k\right)\phi(-z)
\\
 \phantom{\phi(-z)V_-(y)^{-1}}{}
 =V_-(y)^{-1}\exp \left( \sum_{k>0,\,\mathrm{odd}}
\frac{2}{k}\left( \frac{y}{z}\right)^{k}\right)\phi(-z)
 =\frac{1+\frac{y}{z}}{1-\frac{y}{z}}V_-(y)^{-1}\phi(-z).
\end{gather*}
The third formula is proved in a similar way.

We will also use the
following identities which can be found in V.~Kac's book~\cite{KacVAfB}:
\begin{gather*}
(z-y)\partial_y\delta(z-y) =\delta(z-y), \\
(z-y)^{k+1}\partial_y^{k}\delta(z-y) =0, \\
\delta(z-y)a(z) =\delta(z-y)a(y), \\
\partial_y\delta(z-y)a(z) =\partial_y\delta(z-y)\left(a(y)+(z-y)\partial_ya(y)\right).
\end{gather*}

We now calculate
\begin{gather*}
\theta(-z)\theta(y) =V_-(-z)^{-1}\phi(-z)V_+(-z)^{-1}V_-(y)^{-1}\phi(y)V_+(y)^{-1}\\
\hphantom{\theta(-z)\theta(y)}{} =\frac{1-\frac{y}{z}}{1+\frac{y}{z}}
V_-(-z)^{-1}\phi(-z)V_-(y)^{-1}V_+(-z)^{-1}\phi(y)V_+(y)^{-1}\\
\hphantom{\theta(-z)\theta(y)}{}=V_-(-z)^{-1}V_-(y)^{-1}\phi(-z)V_+(-z)^{-1}\phi(y)V_+(y)^{-1}\\
\hphantom{\theta(-z)\theta(y)}{} =\frac{1+\frac{y}{z}}{1-\frac{y}{z}}
V_-(-z)^{-1}V_-(y)^{-1}\phi(-z)\phi(y)V_+(-z)^{-1}V_+(y)^{-1}\\
\hphantom{\theta(-z)\theta(y)}{} =\frac{1+\frac{y}{z}}{1-\frac{y}{z}}
V_-(-z)^{-1}V_-(y)^{-1}\left({:}\phi(-z)\phi(y){:}-\frac{1}{z}\frac{1}{1-\frac{y}{z}}\right)
V_+(-z)^{-1}V_+(y)^{-1}.
\end{gather*}
Now replacing $z$ and $y$ by $-y$ and $-z$ respectively, gives
\[
\theta(y)\theta(-z)=\frac{1+\frac{z}{y}}{1-\frac{z}{y}}
V_-(-z)^{-1}V_-(y)^{-1}\left({:}\phi(-z)\phi(y){:}+\frac{1}{y}\frac{1}{1-\frac{z}{y}}\right)
V_+(-z)^{-1}V_+(y)^{-1}
\]
and thus
 \begin{gather}
 \theta(-z)\theta(y)+\theta(y)\theta(-z)
=
2z\delta(z-y)V_-(-z)^{-1}V_-(y)^{-1}{:}\phi(-z)\phi(y){:}
V_+(-z)^{-1}V_+(y)^{-1}\nonumber\\
\qquad\quad{} + \left( \frac 1y \frac{1+\frac{z}{y}}{(1-\frac{z}{y})^2}-\frac
1z \frac{1+\frac{y}{z}}{(1-\frac{y}{z})^2}\right)
V_-(-z)^{-1}V_-(y)^{-1} V_+(-z)^{-1}V_+(y)^{-1}\nonumber\\
\qquad {}
= 4yH(y)\delta(z-y)-\partial_y\delta(z-y)(y+z)V_-(-z)^{-1}V_-(y)^{-1}
V_+(-z)^{-1}V_+(y)^{-1}\nonumber\\
\qquad {}
= 4yH(y)\delta(z-y)-2y\partial_y\delta(z-y)
-(z-y)\partial_y\delta(z-y) \nonumber\\
\qquad\quad{}
 \times\left(1+2y\partial_y\left(V_-(-y)^{-1}\right)V_-(y)^{-1}
+2y\partial_y\left(V_+(-y)^{-1}\right)V_+(y)^{-1}\right)\nonumber\\
\qquad {}
= 4yH(y)\delta(z-y)-2y\partial_y\delta(z-y)-
\delta(z-y)\left(1+4yH(y)\right)\nonumber\\
\qquad {}
= {-2}y\partial_y\delta(z-y)-\delta(z-y)=-D_y\delta(z-y) ,\label{thetatheta-equality}
\end{gather}
where $H(y)$ is as in \eqref{S4} (not the one in \eqref{Gamma-bt2}),
and $D_y=y\partial_y+\partial_y y$ is the Euler operator.

Now write
\begin{gather}\label{theta(z)}
\theta(z)=2\sum_{i\in\frac{1}{2}+\mathbb{Z}}
J_iz^{-i-\frac{1}{2}} .
\end{gather}
Note that there is no conf\/lict with the $J$'s def\/ined in (\ref{S4}), since here the $J$'s
have indices in $\frac{1}{2}+\mathbb{Z}$.
It is clear that the above commutation
relation \eqref{thetatheta-equality} in modes gives
\[
J_jJ_k+J_kJ_j=(-)^{j-\frac{1}{2}}\frac{j}2 \delta_{j,-k},\qquad j,\,k\in \frac{1}{2}+\mathbb{Z}
\]
(compare with (\ref{S5})). From \eqref{comm-Htheta} we also have
\[
[J_n, J_m]=0 ,\qquad n\in 1+2\mathbb{Z},\quad  m\in \frac{1}{2}+\mathbb{Z} .
\]
Thus we can combine the (anti)commutation relations of all $J$'s as follows:
 \begin{gather}\label{supercurrent}
[J_i, J_j]_s = \frac{j}{2}(-1)^{[j-\frac 12]}\delta_{i,-j} ,
 \end{gather}
where the notation $[\ {,}\ ]_s$ serves for the supercommutator while
$[i]$ denotes the integer part of a~real number~$i$. As we see,
$\deg J_i=-i$
 and that
\[
J_k|0\rangle=\langle 0|J_{-k}=0,\qquad
J_{-k}|0\rangle\ne 0\ne\langle 0|J_{k} \qquad\mbox{for}\ \ k>0 .
\]

\subsection{Even and odd times} \label{Even and odd times}
Since the elements $J_{-k}$ with $k\in \frac{1}{2}+\mathbb{Z}$
anticommute among themselves, they can only
appear once in
\[
J_{-k_n}J_{-k_{n-1}}\cdots J_{-k_3}J_{-k_2}J_{-k_1}|0\rangle.
\]
Such a $J_{-k}$ explains the factor $1+q^k$ in the
$q$-dimension formula \eqref{q3}. One can identify the~$J_n$'s, for $n<0$ with even and odd times, i.e., with  commuting
variables $t_j$, $0<j\in 1+2\mathbb{Z}$, and Grassmann variables $t_{\frac{j}2}$, $0<j\in
1+2\mathbb{Z}$, and identify Fock space $F$ with the space
\[
\mathbb{C}\big[t_{2j-1},t_{\frac{2j-1}2};\, j=1,2,\ldots \big]
\]
(or some completion of it, since we take exponentials),
where one has
\[
t_it_j-(-)^{4ij}t_jt_i=0 ,
\]
in particular, $t_j^2=0$ for $j\in \frac12+\mathbb{Z}$\,.
We will write ${\bf t}=\bigl(t_1,t_{\frac 12};t_3,t_{\frac 32};t_5,t_{\frac
52};\dots \bigr)$ and use $t=\left(t_1,t_3,t_5,\dots \right)$ and
$t_\mathrm{odd}=\bigl(t_{\frac 12},t_{\frac 32};,t_{\frac
52},\dots \bigr)$.

{\it Let $\sigma$ be this isomorphism,
sending $F$ to $\mathbb{C}[t_{2j-1},t_{\frac{2j-1}2}; j=1,2,\ldots]$.}
Then
\begin{gather}\label{theta-xi}
\sigma J_{-j}\sigma^{-1}=(-)^{[\frac{1}{2}-j]}\frac{j}2t_j\qquad\mbox{and}\qquad
\sigma J_j\sigma^{-1}=\frac{\partial}{\partial t_j}\,,\qquad
j>0\,,
\end{gather}
give the f\/ield exactly in commuting and anticommuting variables
$t_k$.

Now using the free boson-(boson+fermion) correspondence, i.e.,
using the vertex operator expressions for the f\/ields
\begin{gather}
\sigma\phi(z)\sigma^{-1} =
\exp\left( \sum_{\scriptstyle0<k\in\mathbb{Z}\atop\scriptstyle k:\,\mathrm{odd}}t_kz^k \right)
\exp\left( \sum_{\scriptstyle0<k\in\mathbb{Z}\atop\scriptstyle k:\,\mathrm{odd}}
\frac{2}{k}\frac{\partial}{\partial t_k}z^{-k} \right)\nonumber\\
\hphantom{\sigma\phi(z)\sigma^{-1} =}{}\times
\sum_{0<j\in\mathbb{Z}}\left( (2j-1)t_{\frac{2j-1}2}(-z)^{j-1}+2\frac{\partial}{\partial
t_{\frac{2j-1}2}}z^{-j} \right)\label{phixi}
\end{gather}
(where we used (\ref{exp}), (\ref{theta}), (\ref{theta(z)}) and (\ref{theta-xi})),
in the following subsections
we shall express the bilinear identity as a hierarchy of dif\/ferential equations.
A similar expression  for~(\ref{phixi}) was also found in~\cite{Kac-Leur}.

\subsection{The CKP bilinear equation}

Following \cite{DJKMVI}  we def\/ine the operator
\[
S=\sum_{k\in\frac{1}{2}+\mathbb{Z}}(-)^{k+\frac{1}{2}}\phi_k\otimes
\phi_{-k} \equiv   \oint \phi(z)\otimes \phi(-z)
\frac{dz}{2\pi i} ,
\]
that $S$ commutes with the action of ${:}\phi_i\phi_j{:}$ on the
tensor product $F\otimes F$ of the Fock space $F$ and
\[
S(|0\rangle \otimes |0\rangle)=
\sum_{k\in\frac{1}{2}+\mathbb{Z}}(-)^{k+\frac{1}{2}}\phi_k|0\rangle
\otimes \phi_{-k}|0\rangle  =0 .
\]

The CKP Hirota equation is \cite{DJKMVI}:
\begin{gather}\label{C-Hirota2}
\oint \phi(z)g|0\rangle \otimes \phi(-z)g|0\rangle\,dz=0 ,
\end{gather}
where  $g$ is for instance given by (see \cite{DJKMVI}):
\begin{gather}
\label{g'} g=\exp\left( -\sum_{m,n}
c_{nm}{:}\phi_{n+\frac{1}{2}}\phi_{m+\frac{1}{2}}{:} \right).
\end{gather}
We rewrite (\ref{C-Hirota2}) as
\begin{gather}\label{C-Hirota3}
\Res_z \phi(z)g|0\rangle \otimes \phi(-z)g|0\rangle =0.
\end{gather}
We could now  use the isomorphism $\sigma$ to def\/ine this hierarchy in terms of the times $\bf t$.
However we will not do that yet, but concentrate f\/irst in the next subsection on the form of $\sigma(g|0\rangle)$.

\subsection{The CKP tau function}

Now let $\rm ODP_{\rm ev}$ be the set of all partitions in an even
number of odd parts, where a part may appear at most once. We call
them ``Odd  Partitions of even length with Distinct parts'' ($\rm
ODP_{\rm ev}$); later on we also need ``Odd Partitions of odd length
with Distinct parts'' ($\rm ODP_{\rm odd}$), their union
$\rm ODP=ODP_{\rm ev}\cup ODP_{\rm odd}$, and the partition 0.
Hence for $0\ne \alpha\in \rm ODP_{\rm ev}$  one has
\[
\alpha=(\alpha_1,\alpha_2,\ldots,\alpha_{2k}) ,
\]
where all $\alpha_i\in 1+2\mathbb{Z}_{\ge0}$, and we assume
\[
\alpha_1>\alpha_2>\cdots>\alpha_{2k} .
\]
Introduce for such partition $0\ne \alpha\in \rm ODP_{\rm ev}$
\[
\xi_\alpha=t_{\frac{\alpha_1}2}
t_{\frac{\alpha_2}2}\cdots t_{\frac{\alpha_{2k}}2} ,\qquad\hbox{and}\qquad
\xi_0=1 .
\]
Then we can rewrite $g|0\rangle$ as
\[
\sigma(g|0\rangle)=\tau({\bf t})=\sum_{\alpha \in \rm ODP_{\rm ev}}
\tau_\alpha(t)\xi_\alpha  .
\]
Note that  $\alpha \in \rm ODP_{\rm ev}$ otherwise $\langle
0|g|0\rangle=0$. We also rewrite  (\ref{C-Hirota3}) as
\[
\Res_z \sum_{\alpha,\beta \in \rm ODP_{\rm
ev}}\sigma\phi(z)\sigma^{-1} \tau_\alpha(t)\xi_\alpha\otimes
\sigma\phi(-z)\sigma^{-1}\tau_\beta(t)\xi_\beta =0 .
\]
It is clear that we can also write
\begin{gather}
\label{prodphi}
\sigma(\phi(z_0)\phi(z_1)\cdots\phi(z_{k})|0\rangle) = f({\bf t})=
\sum_{\alpha\in {\rm ODP},\,\ell(\alpha)\le k+1}
f_\alpha(t)\xi_\alpha .
\end{gather}
Now let
\[
\alpha=(\alpha_1,\alpha_2,\ldots,\alpha_n) ,\qquad
\alpha_i>\alpha_{i+1} .
\]
Then clearly
\[
f_\alpha(t)=\frac{\partial}{\partial t_{\frac{\alpha_n}2}}
\frac{\partial}{\partial t_{\frac{\alpha_{n-1}}2}}\cdots
\frac{\partial}{\partial
t_{\frac{\alpha_1}2}}\sigma(\phi(z_0)\phi(z_1)\cdots\phi(z_{k})|0\rangle)
\biggr|_{t_\mathrm{odd}=0}  .
\]
Now substitute the vertex operator expression for the f\/ields
$\sigma\phi(z_j)\sigma^{-1}$. We thus obtain (assu\-ming
$|z_i|>|z_j|$ if $i<j$):
\[
\sigma(\phi(z_0)\phi(z_1)\cdots\phi(z_{k})|0\rangle)=
\left(\prod_{0\le i<j\le k}\frac{z_i+{z_j}}{z_i-{z_j}}\right)
e^{\sum\limits_{m=0}^k
\varphi(t,z_m)}\sigma(\theta(z_0)\cdots\theta(z_{k})|0\rangle) ,
 \]
where
 \begin{gather}\label{varphi}
\varphi(t,z) = \sum_{0<k\in\mathbb{Z},\,\mathrm{odd}}t_kz^k  .
 \end{gather}
Then setting
$\Pi(z):= \prod\limits_{-1\le i<j\le k}\frac{z_i+{z_j}}{z_i-{z_j}}$
we have
\begin{gather}
\nonumber
f_\alpha(t)  =\Pi(z)
e^{\sum\limits_{m=0}^k \varphi(t,z_m)} \frac{\partial}{\partial
t_{\frac{\alpha_n}2}} \frac{\partial}{\partial t_{\frac{\alpha_{n-1}}2}}
\cdots \frac{\partial}{\partial t_{\frac{\alpha_1}2}}
\sigma\left(\theta(z_0)\cdots\theta(z_{k})|0\rangle\right)\Bigr|_{t_\mathrm{odd}=0}
\\
\phantom{f_\alpha(t)}{}  =\Pi(z)
e^{\sum\limits_{m=0}^k \varphi(t,z_m)} \sigma\left(
J_{\frac{\alpha_n}{2}}J_{\frac{\alpha_{n-1}}{2}} \cdots
J_{\frac{\alpha_{1}}{2}}
\theta(z_0)\cdots\theta(z_{k})|0\rangle\right)\Bigr|_{t_\mathrm{odd}=0}
\nonumber\\
\phantom{f_\alpha(t)}{} =\Pi(z)
e^{\sum\limits_{m=0}^k \varphi(t,z_m)} \langle 0|
J_{\frac{\alpha_n}{2}}J_{\frac{\alpha_{n-1}}{2}} \cdots
J_{\frac{\alpha_{1}}{2}}
\theta(z_0)\cdots\theta(z_{k})|0\rangle
\label{pr1}\\
\phantom{f_\alpha(t)}{} =\frac{\Pi(z)}{2^n}
e^{\sum\limits_{m=0}^k \varphi(t,z_m)}\Res_{z_{-n}}\cdots \Res_{z_{-1}}
z_{-n}^{\frac{\alpha_n}{2}-\frac{1}{2}} \cdots
z_{-1}^{\frac{\alpha_{1}}{2}-\frac{1}{2}} \langle 0|
\theta(z_{-n})
\theta(z_{-n+1})\cdots\theta(z_{k})|0\rangle \nonumber\\
\phantom{f_\alpha(t)}{}=\frac{\Pi(z)}{2^n}
e^{\sum\limits_{m=0}^k \varphi(t,z_m)}\Res_{z_{-n}}\cdots \Res_{z_{-1}}
z_{-n}^{\frac{\alpha_n}{2}-\frac{1}{2}} \cdots
z_{-1}^{\frac{\alpha_{1}}{2}-\frac{1}{2}} \Pf\left( \left(
\frac{z_i-{z_j}}{(z_i+{z_j})^2} \right)_{-n\le i,j\le
k}\right) ,\nonumber
\end{gather}
where Pf stands for the Pfaf\/f\/ian. The last equality follows from
Wick's theorem and from
\[
\langle 0|\theta(y)\theta(z)|0\rangle=\frac{y-{z}}{\left(y+{z}\right)^2}  .
\]

Def\/ine
  \begin{gather}\label{Gamma-bt}
\Gamma({\bf t}):=e^{J({\bf t})}  ,\qquad J({\bf t}) =\sum_{0<
i\in 1+2\Z}  t_i J_{i}+\sum_{0< i\INZ12}  t_i J_{i}  ,
  \end{gather}
where $\deg  t_i = i$, $\deg  J({\bf t}) = 0$.
We write
\begin{gather}\label{Gamma-bt2}
\Gamma({\bf t})=e^{H(t)}e^{\chi(t_\mathrm{odd})} ,\qquad  \mbox{where}\quad
H(t):=\sum_{0< i\in 1+2\Z}  t_i J_{i} ,\qquad
\chi(t_\mathrm{odd}):=\sum_{0< i\INZ12}  t_i J_{i} .
\end{gather}
We will now show that  $f({\bf t})$ is equal to
\[
\langle
0|\Gamma({\bf t})\phi(z_0)\phi(z_1)\cdots\phi(z_{k})|0\rangle  .
\]

Using $e^{H(t)}\phi(z)e^{-H(t)}=e^{\varphi(t,z)}$ and
$ e^{\chi(t_\mathrm{odd})}\phi(z)e^{-\chi(t_\mathrm{odd})}=\phi(z)+\Xi(z)
$,
where
\[
\Xi(z)=V_-(z)\xi(z)V_+(z) , \qquad
\xi(z)=\sum_{0<k\in \frac12+\mathbb{Z}}
k t_k(- z)^{{k}-\frac{1}{2}} ,
\]
we see that
\begin{gather*}
 \frac{\partial}{\partial
t_{\frac{\alpha_n}2}} \frac{\partial}{\partial t_{\frac{\alpha_{n-1}}2}}
\cdots \frac{\partial}{\partial t_{\frac{\alpha_1}2}}\langle 0|
e^{H(t)}e^{\chi(t_\mathrm{odd})}\phi(z_0)\phi(z_1)\cdots\phi(z_{k})
|0\rangle\biggr|_{t_\mathrm{odd}=0}  \\
\qquad{} = \langle
0|e^{H(t)} J_{\frac{\alpha_n}{2}}J_{\frac{\alpha_{n-1}}{2}}
\cdots J_{\frac{\alpha_{1}}{2}}
e^{\chi(t_\mathrm{odd})}\phi(z_0)\phi(z_1)\cdots\phi(z_{k})
|0\rangle\bigr|_{t_\mathrm{odd}=0}\\
\qquad{} =\langle
0|J_{\frac{\alpha_n}{2}}J_{\frac{\alpha_{n-1}}{2}}
\cdots J_{\frac{\alpha_{1}}{2}}
e^{H(t)}e^{\chi(t_\mathrm{odd})}\phi(z_0)\phi(z_1)\cdots\phi(z_{k})
|0\rangle\bigr|_{t_\mathrm{odd}=0}\\
\qquad{} = e^{\sum\limits_{m=0}^k \varphi(t,z_m)}  \langle
0|J_{\frac{\alpha_n}{2}}J_{\frac{\alpha_{n-1}}{2}}
\cdots J_{\frac{\alpha_{1}}{2}}\left(\phi(z_0)+\Xi(z_0)\right)\\
\qquad\quad{}\times
\left(\phi(z_1)+\Xi(z_1)\right)
\cdots\left(\phi(z_{k})+\Xi(z_k)\right)
|0\rangle\bigr|_{t_\mathrm{odd}=0}\\
\qquad{} =e^{\sum\limits_{m=0}^k \varphi(t,z_m)} \langle
0|J_{\frac{\alpha_n}{2}}J_{\frac{\alpha_{n-1}}{2}}
\cdots J_{\frac{\alpha_{1}}{2}} \phi(z_0)\phi(z_1)
\cdots\phi(z_{k})
|0\rangle\\
\qquad{} = e^{\sum\limits_{m=0}^k \varphi(t,z_m)}
 \langle
0|J_{\frac{\alpha_n}{2}}J_{\frac{\alpha_{n-1}}{2}}
\cdots J_{\frac{\alpha_{1}}{2}} \theta(z_{0}) \theta(z_1)
\cdots\theta(z_{k})\\
\qquad\quad{}\times
 V_-(z_0)V_+(z_0)V_-(z_1)V_+(z_1)\cdots
V_-(z_k)V_+(z_k)
|0\rangle\\
\qquad{}  =\left(\prod_{0\le i<j\le k}\frac{z_i+{z_j}}{z_i-{z_j}}\right)
e^{\sum\limits_{m=0}^k \varphi(t,z_m)} \langle
0|J_{\frac{\alpha_n}{2}}J_{\frac{\alpha_{n-1}}{2}}
\cdots J_{\frac{\alpha_{1}}{2}} \theta(z_{0}) \theta(z_1)
\cdots\theta(z_{k})
|0\rangle
 =f_\alpha(t) .
\end{gather*}
Note that from this for $k$ odd, due to Wick's rule for bosons we also have
\begin{gather*}
f_0(t) = e^{\sum\limits_{m=0}^k \varphi(t,z_m)} \langle 0|
\phi(z_0)\phi(z_1) \cdots\phi(z_{k})
|0\rangle =e^{\sum\limits_{m=0}^k \varphi(t,z_m)} \Hf \left(\left(\langle
\phi(z_i)\phi(z_j)\rangle
\right)_{0\le i,j\le k}\right)\\
\phantom{f_0(t)}{} =e^{\sum\limits_{m=0}^k \varphi(t,z_m)}
\Hf\left(\left(\frac{1}{z_i+z_j}\right)_{0\le i,j\le k}\right),
\end{gather*}
where Hf stands for the Hafnian. The Hafnian of  a symmetric
matrix $A$ of even order is def\/ined as follows
\[
\Hf (A) :=\sum_\sigma
 A_{\sigma(1),\sigma(2)}A_{\sigma(3),\sigma(4)}\cdots
A_{\sigma(2k-1),\sigma(2k)} ,
\]
where the sum runs over all permutations $\sigma$ of $\{1,\dots,2k\}$ satisfying
\[
\sigma(2i-1)<\sigma(2i),\qquad\sigma(1)<\sigma(3)<\cdots<\sigma(2k-1).
\]
  As one can see the Hafnian  contains $1\cdot  3\cdot 5\cdot \cdots
  \cdot(2k-1)=:(2k-1)!!$ terms.

\begin{remark}
Comparing this with $f_0$ in (\ref{pr1}), we have a new proof for
the identity
\begin{gather}
\label{pfaffhaf} \Pf\left(\left(
\frac{z_i-{z_j}}{\left(z_i+{z_j}\right)^2} \right)_{1\le i,j\le
2k}\right)= \prod_{1\le i<j\le 2k}\frac{z_i-{z_j}}{z_i+{z_j}}
\Hf\left(\left(\frac{1}{z_i+z_j}\right)_{1\le i,j\le
2k}\right)
\end{gather}
of \cite{IKO}.
\end{remark}

Since $g|0\rangle$, where $g$ is given by (\ref{g'}), is a possibly
inf\/inite linear combination of
\[
\phi_{j_1}\phi_{j_2}\cdots \phi_{j_{2k}}|0\rangle ,
\]
which can be obtained by taking residues of the expression in
(\ref{prodphi}), one deduces that
\begin{gather}
\label{pr7} \sigma(g|0\rangle )=\tau({\bf t})=\langle
0|e^{H(t)}e^{\chi(t_\mathrm{odd})}g|0\rangle = \sum_{\alpha} \tau_\alpha(t)
\xi_{\alpha}
\end{gather}
and
\begin{gather}
\label{pr77} \tau_\alpha(t)=\langle \alpha|e^{H(t)} g|0\rangle ,
\end{gather}
where
\begin{gather}
\label{alpha..} \langle \alpha|=
\langle
0|J_{\frac{\alpha_n}{2}}J_{\frac{\alpha_{n-1}}{2}}
\cdots J_{\frac{\alpha_{1}}{2}} .
\end{gather}

\subsection{A CKP wave function}

Now we want to study $\phi(z)g|0\rangle$. Consider the expression
\begin{gather*}
\sigma(\phi(z)g|0\rangle)= \sigma \phi(z)\sigma^{-1}\sigma(g|0\rangle)
= \sigma \phi(z)\sigma^{-1}\tau({\bf t})\\
\phantom{\sigma(\phi(z)g|0\rangle)}{}
= e^{\varphi(t,z)}
\sum_{0<j\in\mathbb{Z}}\left((2j-1)t_{\frac{2j-1}2}(-z)^{j-1}+2\frac{\partial}{\partial
t_{\frac{2j-1}2}}z^{-j}\right)\times{}
\\
\phantom{\sigma(\phi(z)g|0\rangle)=}{}
\times\exp\left(\sum_{0<k\in\mathbb{Z},\,\mathrm{odd}}\frac{2}{k}\frac{\partial
}{\partial t_k}z^{-k}\right)   \tau({\bf t}) ,
\end{gather*}
where (\ref{phixi}) was used.

Clearly, one also has
\begin{gather}
\label{sphi} \sigma(\phi(z)g|0\rangle)=\langle
0|e^{H(t)}e^{\chi(t_\mathrm{odd})}\phi(z)g|0\rangle .
\end{gather}
We now write $\sigma(\phi(z)g|0\rangle)$ as
\[
\sigma(\phi(z)g|0\rangle)=\sum_{\alpha \in {\rm
ODP}_\mathrm{odd}}g_\alpha(t,z)\xi_\alpha .
\]
Now substitute this in (\ref{C-Hirota3}), omitting the tensor
symbol and writing $s_j$  for $t_j$ in the right-hand side of the tensor product, we obtain,
that for every $\alpha,\beta\in{\rm ODP}_\mathrm{odd}$ the coef\/f\/icient of
$\xi_\alpha\eta_\beta$ is equal to
\begin{gather}
\label{HHir} \Res_zg_\alpha(t,z)g_\beta(s,-z)=0 .
\end{gather}
Now we want to express $g_\alpha(t,z)$ in terms of the
$\tau_\beta(t)$'s.

It will be convenient to introduce some more notation here. Let
\[
\alpha=(\alpha_1,\alpha_2,\ldots,\alpha_{k})\in \rm ODP,
\]
where all $\alpha_i\in 1+2\mathbb{Z}_{\ge0}$, and we assume
\[
\alpha_1>\alpha_2>\cdots>\alpha_{k}.
\]
Now let $\nu\in 1+2\mathbb{Z}_{\ge0}$, $\nu\not\in \alpha$, i.e.,
\[
\nu\not\in\{\alpha_1,\alpha_2,\ldots,\alpha_{k}\},
\]
and
\[
\alpha_i>\nu>\alpha_{i+1} ,
\]
then we def\/ine an ``addition'' as follows
\[
\alpha \cup \nu : =(\alpha_1,\alpha_2,\ldots, \alpha_i, \nu,
\alpha_{i+1},\ldots,\alpha_{k}) .
\]
%Remark.
Note that the notation $\alpha+\beta$ was used dif\/ferently
in \cite{Mac} where it was def\/ined as
$(\alpha_1+\beta_1,\alpha_2+\beta_2,\dots)$.

In a similar way the subtraction $\alpha \backslash \alpha_i$
for $ \alpha_i\in\alpha$ is def\/ined by
\[
\alpha \backslash \alpha_i=(\alpha_1,\alpha_2,\ldots,
\alpha_{i-1}, \alpha_{i+1},\ldots,\alpha_{k}) .
\]
Then for
\[
\alpha=(\alpha_1,\alpha_2,\ldots,\alpha_{k})\in {\rm ODP}_\mathrm{odd} ,
\]
hence $k$ odd, we f\/ind
\begin{gather}
g_\alpha(t,z) =e^{\varphi(t,z)}
\exp\left(\sum_{0<j\in\mathbb{Z},\,\mathrm{odd}}
\frac{2}{j}\frac{\partial}{\partial t_j}z^{-j}\right)\times{}\nonumber\\
\phantom{g_\alpha(t,z) =}{}  \times \left(\sum_{i=1}^{k} (-)^{i-1}
\alpha_i\tau_{\alpha\backslash \alpha_i}(t)(-z)^{\frac{\alpha_i
-1}{2}} + 2\!\sum_{\nu\in 1+ 2\mathbb{Z}_{\ge0},\, \nu\not\in \alpha}\!
s(\nu,\alpha)\ \tau_{\alpha\cup \nu}(t)z^{-\frac{\nu
+1}{2}}\right) ,\!\!\label{Wave}
\end{gather}
where
\[
s(\nu,\alpha)=(-1)^{|\{ \alpha_i\in \alpha|\, \alpha_i>\nu\} |} .
\]
In particular for $\alpha=(1)=1$ we f\/ind
\[
g_{1}(t,z)= e^{\varphi(t,z)}
\exp\left(\sum_{0<j\in\mathbb{Z},\,\mathrm{odd}}
\frac{2}{j}\frac{\partial}{\partial
t_j}z^{-j}\right)\left(\tau_0(t) +2 \sum_{\nu\in 1+
2\mathbb{Z},\, \nu>1} \tau_{(\nu,1)}(t)z^{-\frac{\nu +1}{2}}\right) ,
\]

We now want to calculate $g_\alpha(t,z)$ as some expectation
value. Using (\ref{sphi}), we have
\begin{gather}
g_\alpha(t,z) = \frac{\partial}{\partial t_{\frac{\alpha_n}2}}
\frac{\partial}{\partial t_{\frac{\alpha_{n-1}}2}} \cdots
\frac{\partial}{\partial t_{\frac{\alpha_1}2}}
\langle 0|e^{H(t)}e^{\chi(t_\mathrm{odd})} \phi(z)g|0\rangle\biggr|_{t_\mathrm{odd}=0}\nonumber\\
\phantom{g_\alpha(t,z)}{}  = \frac1{2^n}\langle
0|\theta_{\frac{\alpha_n}{2}}\theta_{\frac{\alpha_{n-1}}{2}}
\cdots
\theta_{\frac{\alpha_{1}}{2}}e^{H(t)} \phi(z)g|0\rangle  =\langle \alpha|e^{H(t)} \phi(z)g|0\rangle  .\label{glam}
\end{gather}
where we have used (\ref{alpha..}).

Now concentrating on (\ref{Wave}) and
 divide this by
 $(-)^{\frac{\alpha_1-1}{2}}\alpha_1\tau_{\alpha\backslash\alpha_1}(t)$, this
 gives
\begin{gather}
w_\alpha(t,z) ={\hat
w}_\alpha(t,z)z^{\frac{\alpha_1-1}{2}}e^{\varphi(t,z)}, \qquad
\mbox{where}\nonumber\\
{\hat w}_\alpha(t,z) =
(-z)^{-\frac{\alpha_1-1}{2}}\frac{g_{\alpha}(t,z)e^{-\varphi(t,z)} }
{\alpha_1\tau_{\alpha\backslash\alpha_1}(t)} =1+O\big(z^{-1}\big),\label{wave}
\end{gather}
which we call the wave function corresponding to $\alpha\in {\rm ODP}$.
Now using (\ref{pr77}) and  (\ref{glam}), one also has
\[
w_\alpha(t,z)=\frac{(-)^{\frac{\alpha_1-1}{2}}}{\alpha_1}
\frac{\langle \alpha|e^{H(t)} \phi(z)g|0\rangle}{\langle
\alpha\backslash \alpha_1|g|0\rangle}.
\]

\subsection{A bilinear identity for the wave function and a Lax equation}

Using the def\/inition of  the wave function (\ref{wave}) we can rewrite   (\ref{HHir}) into
a bilinear identity for these wave functions:
\begin{gather}
\Res_z w_\alpha(t,z)w_\beta(s,-z)\nonumber\\
\qquad{}=\Res_z {\hat
w}_\alpha(t,z){\hat w}_\beta(s,-z)
e^{\varphi(t,z)-\varphi(s,z)}z^{\frac{\alpha_1+\beta_1-2}{2}}=0,
\qquad \alpha,\beta\in {\rm ODP}_\mathrm{odd}.\label{Hirwave}
\end{gather}

Let us take $\alpha=\beta$ and $\alpha_1=\beta_1=n$, $n$ odd. Then equation \eqref{Hirwave} takes the form
 \[
  \Res_z {\hat
w}_\alpha(t,z){\hat w}_\alpha(s,-z)
e^{\varphi(t,z)-\varphi(s,z)}z^{n-1}=0 ,
\qquad  \alpha_1=\beta_1=n ,\quad  n \  \mbox{odd} ,
 \]
which coincides with equations (5) and $(12)_n$ ($n$ odd) of \cite{DJKMVI} where $w(x,k)$ should be compared with our
$e^{\varphi(t,z)}{w}_\alpha(t,z)$.

Now write for $\alpha=(1)=1$
\[
w(t,z)=w_1(t,z),\qquad\mbox{and}\qquad {\hat w}(t,z)={\hat
w}_1(t,z)
\]
then we can see this as the Date, Jimbo, Kashiwara, Miwa CKP wave
function as in \cite{DJKMVI},
 since
in particular
\begin{gather}
\label{HHirwave} \Res_z {\hat w}(t,z){\hat w}(s,-z)
e^{\varphi(t,z)-\varphi(s,z)}=0.
\end{gather}
We can rewrite the wave functions as follows
\[
w_\alpha(t,z)={\hat
w}_\alpha(t,\partial)\partial^{\frac{\alpha_1-1}{2}}e^{\varphi(t,z)},
\]
where $\partial =\frac{\partial}{\partial x}=\frac{\partial}{\partial t_1}$. Then ${\hat
w}_\alpha(t,\partial)$ is a pseudo dif\/ferential operator of order~0. Note that
\[
w_\alpha(t,-z)={\hat w}_\alpha
(t,\partial)\partial^{\frac{\alpha_1-1}{2}}e^{-\varphi(t,z)}.
\]

Now we use the following known lemma (see, e.g., \cite{KL}):

\begin{lemma} For pseudodifferential operators $P(t,\partial)$
and
 $Q(t,\partial)$ we have
\begin{gather*}
\Res_z \left( P(t,\partial)\cdot
\exp\left(\sum_{0<j\in\mathbb{Z},\,\mathrm{odd}}t_jz^j\right)\right)
\left( Q(t,\partial)\cdot \exp\left(-\sum_{0<j\in\mathbb{Z},
\,\mathrm{odd}}t_jz^j\right)\right)
\\
\qquad {}=\Res_\partial P(t,\partial)\cdot
 Q(t,\partial)^*=0.
\end{gather*}
where the conjugation and $\Res_\partial$ are defined on
monomials respectively as
$\left(a(x)\partial^k\right)^*=(-\partial)^k a(x)$ and
$\Res_\partial a(x)\partial^k=a(x)\delta_{k,-1}$.
 \end{lemma}

Taking $P={\hat w}$ and $Q=\partial^n{\hat w}$, $n=0,1,2,\dots$,
by~(\ref{HHirwave}) we obtain that
 \begin{gather}\label{alpha-beta}
\left({\hat
w}_\alpha(t,\partial)\partial^{\frac{\alpha_1+\beta_1-2}{2}}{\hat
w}_\beta(t,\partial)^*\right)_-=0 .
 \end{gather}
Now we take $\alpha=\beta=(1)=1$. Then, since ${\hat
w}(t,\partial)= 1+\sum\limits_{k=1}^\infty w_k(t)\partial^{-k}$, one
obtains that
\begin{gather}
\label{inv-wave} {\hat w}(t,\partial)^*={\hat w}(t,\partial)^{-1} .
\end{gather}
Start again with (\ref{HHirwave}) and dif\/ferentiate this equation
in $t_k$ for $k$ odd; then one gets
\[
\Res_z \left(\frac{\partial {\hat w}(t,z)}{\partial
t_k}+{\hat w}(t,z)z^k\right) \exp\left(\sum_{0<j\in\mathbb{Z},\,
{\rm odd}}t_jz^j\right)w(s,-z)=0 .
\]
Again using the fundamental Lemma and (\ref{inv-wave}), we deduce
\[
\left(\left(\frac{\partial {\hat w}(t,\partial)}{\partial t_k}
+{\hat
w}(t,\partial)\partial^k\right)W(t,\partial)^{-1}\right)_-=0 ,
\]
which gives the Sato--Wilson equation:
\[
\frac{\partial {\hat w}(t,\partial)}{\partial t_k} =-\left({\hat
w}(t,\partial)\partial^k{\hat w}(t,\partial)^{-1}\right)_- {\hat
w}(t,\partial) .
\]

This is equivalent to the following equation for the wave function
\[
\frac{\partial w(t,z)}{\partial t_k}= \left({\hat
w}(t,\partial)\partial^k{\hat w}(t,\partial)^{-1}\right)_+w(t,z).
\]
Introducing the Lax operator
\[
L(t,\partial)={\hat w}(t,\partial)\partial {\hat
w}(t,\partial)^{-1} =\partial+\sum_{k=1}^\infty
u_k(t)\partial^{-k},
\]
then (\ref{inv-wave}) induces $L^*=-L$ and
from the Sato--Wilson equation one derives the following CKP Lax equation:
\[
\frac{\partial L(t,\partial)}{\partial t_k} =\Big[ \big(
L(t,\partial)^k\big)_+,L(t,\partial)\Big].
\]

\section{Bilinear identity in super notations} % \label{super}

In the previous section we obtained a wave function. In this section our approach will be slightly dif\/ferent.
We want to superize, i.e., obtain a supersymmetric wave function that also include the Grassmannian times
$t_i$ with $i\in\frac12+\mathbb{Z}$ and the corresponding bilinear equation for this super
 wave function~(\ref{super-bil}). Let us point out that in this way we shall re-write results of the previous
section using super notations. We regard this an important step, which might be very fundamental for the further development of the theory. However, unfortunately we were not able to obtain Lax equations with respect to odd Grassmannian times in this setting.

\subsection{Super vertex operator}
Recall the super  commutation relations (\ref{supercurrent}) and the super times
(\ref{supercurrent}) and the def\/inition of~$\Gamma({\bf t})$ in (\ref{Gamma-bt}).
It is convenient to introduce an auxiliary parameter $\zeta$, which
is a Grassmannian variable, an odd counterpart to $z$:
$\zeta^2=0$, $z\zeta=\zeta z$, $\deg z = 2\deg\zeta=-1$.

Introduce the following ``superfermionic'' f\/ields
\begin{gather*}
\Phi(z,\zeta)  :=  2\sum_{n\in\Z} \zeta
 \frac{z^{-2n-1}}{2n+1}J_{2n+1}-2\sum_{0\neq
n\in\Z}\frac{z^{-n}}{n}J_{n+\frac 12}  + 2 J_{\frac12}\log z ,
\\
\Theta(z,\zeta)  :=  \left(\frac{\partial}{\partial
\zeta}+\zeta\frac{\partial}{\partial z} \right)\Phi(z,\zeta) .
\end{gather*}
 Here $\frac{\partial}{\partial
\zeta}+\zeta\frac{\partial}{\partial z} $ is a superderivative
which will be denoted by $D_{z,\zeta}$,
$D_{z,\zeta}^{\ 2}=\frac{\partial}{\partial z}$. As we see $\deg
\Phi =-\frac 12$ while $\deg \Theta=0$.

Let $ {:}e^{\Theta(z,\zeta)}{:}$ denote $
e^{\Theta_-(z,\zeta)}e^{\Theta_+(z,\zeta)}$ where $\Theta_\pm$
denotes the splitting of $\Theta$ in series in respectively
positive/negative powers of $z^{-1}$. One can verify
\begin{gather*}
{:}e^{\Theta(z_1,\zeta_1)}{:}\,{:}e^{\Theta(z_2,\zeta_2)}{:}=
 e^{\Theta_-(z_1,\zeta_1)+\Theta_-(z_2,\zeta_2)}e^{\Theta_+(z_1,\zeta_1)+\Theta_+
 (z_2,\zeta_2)}
\left(\frac{z_1+z_2}{z_1-z_2}+\frac{\zeta_1\zeta_2}{z_1+z_2}
\right) .
\end{gather*}
 It follows from $\zeta^2=0$ that
  \begin{gather} \label{super-vertex-viaVphi}
  {:}e^{\Theta(z,\zeta)}{:}= V_-(z)V_+(z)+\zeta\phi(z) .
  \end{gather}
Then
 \begin{gather}
\phi(z)  =  \frac{\partial}{\partial \zeta}\,
{:}e^{\Theta(z,\zeta)}{:}\nonumber\\
\hphantom{\phi(z)}{}
=\frac{\partial}{\partial
\zeta}\left(e^{\sum\limits_{0>n\in\Z}\left(\frac{2z^{-2n-1}}{2n+1}J_{2n+1}+
2\zeta z^{-1-n}J_{n+\frac 12}\right)}\cdot
e^{\sum\limits_{0\le n\in\Z}\left(\frac{2z^{-2n-1}}{2n+1}J_{2n+1} +
2\zeta z^{-n-1}J_{n+\frac 12}\right)}\right).\!\!\!\label{phi-Theta}
\end{gather}
Using the isomorphism $\sigma$, see (\ref{theta-xi}), we can express
\begin{gather}\label{varphi-potential}
\sigma \Theta_-(z,\zeta)\sigma^{-1}  =
\sum_{0<i\in 1+2\Z}  z^{i}t_i + \zeta
\sum_{0<i\INZ12}i(-z)^{i-\frac12}t_i =: \varphi({\bf
t},z,\zeta) ,
\end{gather}
and
\[
 \sigma \Theta_+(z,\zeta)\sigma^{-1}
= 2\sum_{0<n\in\Z} \frac{z^{1-2n}}{2n-1}\frac{\partial}{\partial
t_{2n-1}} + 2\zeta
\sum_{0<n\in\Z}{z^{-n}}\frac{\partial}{\partial t_{n-\frac 12}} .
\]
Now
 \begin{gather}\label{Bose-vertex}
\sigma \phi(z)
\sigma^{-1}=\frac{\partial}{\partial\zeta}\Upsilon({\bf
t},z,\zeta) ,
 \end{gather}
where
\begin{gather}
 \Upsilon({\bf t},z,\zeta):=e^{\varphi({\bf t},z,\zeta)} \Upsilon_+(z,\zeta) ,
\nonumber\\
\Upsilon_+(z,\zeta)=\exp \left(2\sum_{0<i\in
1+2\Z}\,\frac{z^{-i}}{i}\frac{\partial}{\partial t_i} + 2\zeta
\sum_{0<i\INZ12}\,{z^{-i-\frac12}}\frac{\partial}{\partial
t_i}\right) .\label{Upsilon+}
 \end{gather}

\subsection{Tau function and super wave function}

Recall (\ref{pr7}), (\ref{pr77}) the tau function as a function of a collection of
both even and odd time variables as follows
  \begin{gather}\label{superTau}
\tau({\bf t}):=\langle 0|\,\Gamma({\bf t})  g \,|0\rangle =
\sum_\alpha  \tau_\alpha(t) \xi_\alpha .
  \end{gather}
As we can see
  \[
\tau({\bf t})=\tau^{{\rm CKP}}( t)+\cdots=\tau_0( t)+\cdots ,
  \]
  where the dots mean terms vanishing when we substitute $t_{\frac12}= t_{\frac32}=\cdots =0$.
  We put
 \begin{gather}\label{super-wave}
 W({\bf t},z,\zeta):=\frac{1}{\tau({\bf t})}\langle 0|\,\Gamma({\bf t})\,
 {:}e^{\Theta(z,\zeta)}{:} \, g
 \,|0\rangle
\end{gather}
  to be the wave function.

From (\ref{Bose-vertex}) we can write
 \begin{gather}\label{Wtau}
 W({\bf t},z,\zeta):=\frac{\Upsilon({\bf t},z,\zeta)\tau({\bf t})}{\tau({\bf
 t})} = e^{\varphi({\bf t},z,\zeta)}\left( 1+O\left(\frac 1z \right)\right) ,
  \end{gather}
where $\varphi({\bf t},z,\zeta)$ was def\/ined in \eqref{varphi-potential}. Let us mark that if $t_\mathrm{odd}=0$ then
$\varphi({\bf t},z,\zeta)$ coincides with $\varphi({ t},z,\zeta)$ def\/ined in \eqref{varphi}.

Note that since $\phi(z)=\frac{\partial}{\partial
\zeta}\,{:}e^{\Theta(z,\zeta)}{:}$ it follows from (\ref{Wave}) and  (\ref{glam}) that for
$\alpha=(\alpha_1,\dots,\alpha_n)$
\[
g_\alpha({t},z) = \left(\frac{\partial}{\partial
\zeta}\frac{\partial}{\partial t_{\frac{\alpha_{n}}{2}}}\cdots
\frac{\partial}{\partial t_{\frac{\alpha_{1}}{2}}} \Upsilon({\bf
t},\zeta,z)\tau({\bf t}) \right)\biggr|_{t_\mathrm{odd} =0},
\]
 where we remind that in our notations: $ t=\left(t_1,t_3,t_5,\dots \right)$ and
 ${\bf t}=
 \left(t_1,t_{\frac 12};t_3,t_{\frac 32};t_5,t_{\frac 52};\dots \right)$.

\subsection{Super Miwa variables}

Denote ${\bf z}=(z_1,\zeta_1;\ldots;z_k,\zeta_k)$, where
$\zeta_i$ are Grassmannian odd and $z_i$ are Grassmannian even
variables, $\zeta_i^2=0$,  and $\deg z_i = -1$, $\deg \zeta_i = -\frac
12$. Let $ [{\bf z}]:=\left(t_1,t_{\frac 12};t_3,t_{\frac 32};t_5,t_{\frac 52};\dots
 \right)$, where
 \begin{gather}\label{s-Miwa}
t_{2n+1}=\frac{2}{(2n+1)}\sum_{i=1}^k\frac 1{z_i^{2n+1}} ,\qquad
t_{n+\frac12} = 2\sum_{i=1}^k\frac{\zeta_i}{z_i^{n}} .
  \end{gather}
  For $k=1$ we shall write $[{\bf z}]$ as $[z,\zeta ]$.
Then
\[
W({\bf t},z,\zeta):=e^{\varphi({\bf t},z,\zeta)}\frac{\tau({\bf
t}-[{z},\zeta])}{\tau({\bf t})}.
\]

In general
\[
\langle 0|\,\Gamma\left( {\bf t}-[{\bf z}]\right)=
\texttt{D}({\bf z})^{-1}\,\langle 0|\,\prod_{i=1}^k {:}
e^{\Theta(z_i,\zeta_i)}{:}\,\Gamma( {\bf t}) ,
\]
where
\[
\texttt{D}({\bf z}) = \texttt{D}_k({\bf z}) := \l
0|\,{:}e^{\Theta(z_1,\zeta_1)}{:}\, \cdots
\,{:}e^{\Theta(z_k,\zeta_k)}{:}\,|0\r,
\]
 which is antisymmetric function with respect to the permutation
 of pairs $(z_i,\zeta_i)$. For instance
  \[
\texttt{D}_2({\bf
z})=\frac{z_1+z_2}{z_1-z_2}+\frac{\zeta_1\zeta_2}{z_1+z_2}.
  \]
We have
 \begin{gather}\label{D^-1}
\texttt{D}_k({\bf
z})^{-1}= \left(\prod_{a<b}\frac{z_a-z_b}{z_a+z_b}\right) \sum_{n=0}^{\left[
k/2\right]}(-1)^{n}\sum_{\alpha\in{\rm ODP}_{\rm ev}\atop
\ell(\alpha)=2n} \zeta_\alpha \Pf\left[ \frac
{z_{\alpha_i}-z_{\alpha_j}}{(z_{\alpha_i}+z_{\alpha_j})^2} \right].
 \end{gather}

\subsection{Bilinear identity for the super wave function $W$}

  The bilinear equation (\ref{HHir}),
  \[
\Res_zg_\alpha({t'},z)g_\beta({t},-z) =0 ,
 \]
  may be written in superized form
 \begin{gather}\label{super-bil}
\Ber_{z,\zeta} \left(\left(\frac{\partial}{\partial
\zeta}W({\bf t'},z,\zeta)\right)W({\bf t},-z,\mp \zeta) \right)
 = 0 ,
\end{gather}
  where $\Ber_{z,\zeta}\,f(z,\zeta):=\Res_z
  \frac{\partial f(z,\zeta)}{\partial
  \zeta}$. The validity of (\ref{super-bil}) follows directly from
  (\ref{C-Hirota3}), (\ref{super-wave}), (\ref{phi-Theta}) and the def\/inition of the Berezinian
  $\Ber$.

\subsection{Bilinear identity as identities for super PDOs}

Let us re-write super-bilinear identities (\ref{super-bil}) in
form of identities for (super version of) pseudo-dif\/ferential
operators (PDO). We will do it in a way similar to the KP case
where PDO operators naturally appear in the framework of KP
bilinear identity \cite{DJKM}.

Now notice that (\ref{Wtau}) yields
 \begin{gather}\label{WW1W2}
 W({\bf t},z,\zeta) =:
  {\hat W}({\bf t},z,\zeta)e^{\varphi({\bf t},z,\zeta)}=:
 \left({\hat W}^{(0)}({\bf t},z)+\zeta{\hat W}^{(1)}({\bf
t},z)\right)e^{\varphi({\bf t},z,\zeta)} ,
  \end{gather}
where (\ref{Upsilon+}) provides
\[
 {\hat W}^{(1)}({\bf t},z)=\sum_{n > 0}{\hat W}_{n}^{(1)}({\bf t})z^{-n}=
 O\left(\frac 1z \right),\qquad
{\hat W}^{(0)}({\bf t},z)=\sum_{n\ge 0}{\hat W}_{n}^{(0)}({\bf
t})z^{-n}=1+O\left(\frac 1z \right) .
\]

For a simplif\/ication we shall denote $x=t_1$ and $\xi=\frac 12
t_{\frac 12}$, thus $\varphi({\bf t},z,\zeta)=zx+\zeta\xi+\cdots$,
and we have
 \[
\partial_x e^{\varphi({\bf t},z,\zeta)} = z e^{\varphi({\bf
t},z,\zeta)} ,\qquad \partial_{\xi} e^{\varphi({\bf
t},z,\zeta)} = - \zeta e^{\varphi({\bf t},z,\zeta)} .
 \]
   Let
 $D := D_{x,\xi}=\partial_\xi+\xi\partial_x$, $D^2=\partial_x$. Then $\partial_\xi
 =D-\xi D^2$. Below we consider the action of odd negative powers of $D$ on
the exponentials $e^{zx+\zeta\xi}$, provided we def\/ine $\partial^n
e^{zx}:=z^ne^{zx}$, $n\in\Z$. To do this we write
$D^{1-2n}=D\partial^{-n}$.
 In such a way we write
  \begin{gather}
D^{2n}\cdot e^{\varphi({\bf t},z,\zeta)} = z^n e^{\varphi({\bf
t},z,\zeta)}, \!\qquad D^{2n+1}\cdot e^{\varphi({\bf
t},z,\zeta)} = z^n\big({-}\zeta + \xi D^2\big) e^{\varphi({\bf
t},z,\zeta)},\!\qquad n\in\Z ;\label{negative-powers}\!\!\!
  \end{gather}
  in particular
  \begin{gather}\label{eta-D}
\zeta  e^{\varphi({\bf t},z,\zeta)}=-\left(D-\xi D^2 \right)\cdot
e^{\varphi({\bf t},z,\zeta)} .
  \end{gather}

Let us introduce
\begin{gather}
K({\bf t},D):=\sum_{n\ge 0} K_n({\bf t}) D^{-n}
:= {\hat W}^{(0)}({\bf t},\partial) - {\hat W}^{(1)}({\bf
 t},\partial)\frac{\partial}{\partial \xi}\nonumber\\
\hphantom{K({\bf t},D)}{} = {\hat W}^{(0)}({\bf t},D^2) - {\hat W}^{(1)}({\bf
 t},D^2)\left(D-\xi D^2 \right)\nonumber\\
\hphantom{K({\bf t},D)}{}
= 1+  {\hat W}_{1}^{(1)}({\bf
 t}) \xi
  +\sum_{n\ge
1}\left({\hat W}_{n}^{(0)}({\bf t})+ {\hat W}_{n+1}^{(1)}({\bf
t})\xi \right)D^{-2n} -\sum_{n \ge 1}{\hat W}_{n}^{(1)}({\bf
t})D^{1-2n} ,\label{W(D)}
\end{gather}
 where negative powers of $D$ are to be understood in the sense of~(\ref{negative-powers}).

From  (\ref{eta-D}), (\ref{W(D)}) and (\ref{WW1W2}) it follows that
\[
 W({\bf t},z,\zeta) =
  K({\bf t},D)\cdot e^{\varphi({\bf t},z,\zeta)} .
\]

In the Lemma below star means the conjugation in the algebra of super PDOs
with properties $(ab)^*=\pm b^*a^*$, where $-$ is
taken if\/f both $a$ and $b$ are odd. We have
$(\partial_x)^*=-\partial_x,\, (\partial_\xi)^*=-\partial_\xi$.
For instance $D^*=-\partial_\xi+(-\partial_x)\xi=-D$,
$\left(D^2\right)^*=-D^*D^*=-\partial$,
$(\xi\zeta)^*=-\zeta\xi=\xi\zeta$, and $(a
D^n)^*=(-1)^{\left[n/2\right]} D^n a$ if $a$ is odd.

We def\/ine $\Res_D\sum\limits_{n\in\Z}
f_n(x,\xi)D^n :=f_{-1}(x,\xi)$. Notice that
  \begin{gather}\label{res-property}
  \Res_D
f(x,\xi)D^{-1}g(x,\xi) =  \pm \Res_D
f(x,\xi)g(x,\xi)D^{-1} ,
 \end{gather}
 where $+$ and $-$ are taken if $g$ is
respectively even and odd.

 \begin{lemma} \label{SuperLemma} Let $P(D)=\sum_n P_n(x,\xi) D^n$ be an odd and
 $Q(D)=\sum_n Q_n(x,\xi) D^n$ an even PDO. Then
\begin{gather}\label{super-Lemma}
\Ber_{z,\zeta}  \big(P(D)\cdot e^{zx+\zeta\xi}\big)\big(Q(D)\cdot
e^{-zx-\zeta\xi}\big) = \Res_D P(D)\left(Q(D)\right)^* ,
\\
\label{super-Lemma-2}
\Ber_{z,\zeta}  \big(P(D)\cdot e^{zx+\zeta\xi}\big)\big(Q(D)\cdot
e^{-zx+\zeta\xi}\big) = \Res_D P(D)\left(Q(-D)\right)^* .
\end{gather}
 \end{lemma}

\begin{proof} Let
 \[
P(D)=\sum_{n\in\Z}\left(
P_n^{(0)}D^{2n}+P_n^{(1)}D^{2n}D^{-1}\right) ,\qquad
Q(D)=\sum_{n\in\Z}\left(
Q_n^{(0)}D^{2n}+Q_n^{(1)}D^{2n}D^{-1}\right) ,
 \]
where $P_n^{(0)}$, $Q_n^{(1)}$  are odd and $P_n^{(1)}$, $Q_n^{(0)}$ are
even. Then{\samepage
 \[
(Q(D))^* = \sum_{n\in\Z}\left(
(-1)^nD^{2n}Q_n^{(0)}-(-1)^nD^{-1}D^{2n}Q_n^{(1)}\right) ,
 \]
where we used
$
(D^*)^{2n}=(-1)^nD^{2n}$, $\left(
D^{-1}\right)^{*}=\left( D^{-2}D\right)^{*}=\left( D\right)^{-1}
$.  }

Using (\ref{res-property}), we f\/ind that the right-hand side
of (\ref{super-Lemma}) is equal to
  \begin{gather}\label{Lemmaterms}
\sum_{n\in\Z}(-1)^n\left(P_n^{(1)}Q_{-n}^{(0)}+
P_n^{(0)}Q_{-n}^{(1)} \right) .
  \end{gather}

 Next consider the left-hand side of (\ref{super-Lemma}). We have
\begin{gather*}
P(D)\cdot e^{zx+\zeta\xi} =\sum_{n\in\Z}\left(
P_n^{(0)}z^{n}+P_n^{(1)}z^{n-1}\left(\zeta+z\xi
\right)\right)e^{zx+\zeta\xi} ,
\\
Q(D)\cdot e^{-zx-\zeta\xi} =\sum_{n\in\Z}\left(
Q_n^{(0)}(-z)^{n}-Q_n^{(1)}(-z)^{n-1}\left(\zeta+z\xi
\right)\right)e^{-zx-\zeta\xi} .
\end{gather*}
   The evaluation of the $\Ber$ of the product of these
   two results in  (\ref{Lemmaterms}).

A similar calculation yields (\ref{super-Lemma-2}).
\end{proof}

 We re-write (\ref{super-bil}) as
\[
\Ber_{z,\zeta} \left(K({\bf
t'},D)\cdot\frac{\partial}{\partial \zeta} e^{\varphi({\bf
t'},z,\zeta)}\right) \left(K({\bf t},D)\cdot e^{\varphi({\bf
t},-z,\mp \zeta)}\right)  = 0.
\]
Taking into account
  \[
\frac{\partial}{\partial \zeta} e^{\varphi({\bf
t},z,\zeta)} = -  \sum_{n\ge 0} \left( (-1)^n\left(n+\frac
1{2}\right)t_{n+\frac 12}\partial^{n} \right)\cdot e^{\varphi({\bf
t},z,\zeta)},
  \]
 where $\partial:=\frac{\partial}{\partial t_1}=D^2$, and
 $\varphi({\bf t},z,\zeta)$ is as in \eqref{varphi-potential},
we obtain
\begin{gather*}
  \Res_D \Bigg( K({\bf t'},D)\cdot
\left(\sum_{n\ge 0} (-1)^n\left(  n+\frac12 \right)t_{n+\frac
12}'D^{2n} \right) \\
\qquad {}\times e^{\pm\left(D-\xi D^2\right)
\sum\limits_{n\ge 0}  (n+\frac 12)t_{n+\frac12}D^{2n} } \cdot K^*({\bf t},\pm D)
\Bigg)=0,
\end{gather*}
which results in
\begin{gather*}
   \left(K({\bf t},D)\cdot
\left(\sum_{n\ge 0} (-1)^n\!\left(n+\frac 1{2}\right)t_{n+\frac
12}D^{2n} \right)  e^{\pm\left(D-\xi D^2\!\right)
\!\sum\limits_{n\ge 0} \! (n+\frac 12)t_{n+\frac12}D^{2n} }\! \cdot K^*({\bf t},\pm D)\right)_-  = 0,
\end{gather*}
  where the subscript $-$ means the taking of projection on series
  with negative powers. Thus
\begin{gather*}
L_\pm := K({\bf t},D)\cdot \left(\sum_{n\ge
0} (-1)^n\left(n+\frac 1{2}\right)t_{n+\frac 12}D^{2n} \right)  e^{\pm\left(D-\xi D^2\right)
\sum\limits_{n\ge 0}  (n+\frac 12)t_{n+\frac12}D^{2n} }
\cdot K^*({\bf t},\pm D)
\end{gather*}
are dif\/ferential operators. These equations are basically equivalent to the set of equations \eqref{alpha-beta}.

\section{Related symmetric functions\label{Related}}

In this section we want to introduce polynomial functions, $\C\lambda$, related to the basis vectors $|\lambda \r $ of
the bosonic Fock space $F$ as the image of the mapping $\sigma$ described in the Subsection~\ref{Even and odd times}. These are
polynomials in Grassmannian even and odd variables $\bt$. In super Miwa variables $z_i$, $\zeta_i$, $i=1,\dots, k$ these
functions are symmetric with respect to the action of the permutation group $S_k$ on the set of pairs $(z_i,\zeta_i)$,
and polynomial with respect to variables $\zeta_i$ and $x_i:=z_i^{-1}$.

These polynomials may be considered as analogues of the celebrated Schur (and projective Schur) functions which are related to
Fock space of charged (resp.\ neutral) fermions. The theoretic f\/ield construction of new functions allows to derive certain
properties which are similar to the properties of the Schur and the projective Schur functions.

\subsection[Polynomials $\C\lambda$]{Polynomials $\boldsymbol{\C\lambda}$}

Let us introduce  suitable notations for the basis of bosonic Fock
vectors in~$F$ (see (\ref{S2BASIS})) and in~$F^*$ labeled by
partitions whose parts are odd numbers
 \begin{gather} \label{basis-vectors} |\lambda\r
:=\frac{1}{{d_\lambda}}\phi_{\frac{\lambda_1}2}\cdots
\phi_{\frac{\lambda_k}2} |0\r ,\qquad \l \lambda | :=\frac{
1}{d_\lambda}\l 0|\phi_{-\frac{\lambda_k}2} \cdots
\phi_{-\frac{\lambda_1}2} ,
 \end{gather}
  where $\lambda=(\lambda_1,\lambda_2,\dots,\lambda_k)$ is a set
  of odd numbers and
   $\lambda_1 \ge\lambda_2\ge\dots\ge\lambda_k>0$, $
   \ell(\lambda):=
k=1,2,\dots  $. The set of partitions with odd parts will be
denoted by OP.

{\it Note that the above vectors     $\l \lambda |$ differ from the  vector $\l \alpha |$ as defined in~\eqref{alpha..},
that is the reason why we write $\lambda$ here to avoid this confusion.}

 In this section we shall use, besides the parts $\lambda_i$ of partitions~$\lambda$, also
the variables $n_i=0,1,2,\dots$ related to odd numbers $\lambda_i$
as follows
 \begin{gather}\label{n-lambda}
\lambda_i =:   2n_i+1 .
 \end{gather}
The sum $|\lambda|:=\lambda_1+\cdots + \lambda_k$ is called the
weight of the partition $\lambda$.  Here the factor $d_\lambda$
for $\lambda\in \rm OP$ is def\/ined by
\begin{gather*}
(d_\lambda)^2:= (-1)^{\sum\limits_{i=1}^k n_i} \!\prod_{i= 1,2,3,\dots}\!
m_{2i-1}!= \prod_{i= 1,2,3,\dots}\! m_{2i-1}!{(-1)^{
m_{4i-1}}} = (-1)^{\frac 12(|\lambda|-\ell(\lambda))} \!\prod_{i=
1,3,5,\dots}\! m_i!,
\end{gather*}
 where $m_i=m_i(\lambda)$ is the number of parts
of $\lambda$ equal to $i$ (or, the same, the multiplicity of~$i$).
(Then we can denote the partition by its frequency notation
$\lambda=(1^{m_1}3^{m_3}5^{m_5}\cdots)$). For instance $
d_{(1^n3^m5^k)}=n!m!(-1)^{m}k! $.

From (\ref{S1}),
 \[
\phi_{-\frac 12\lambda_i}\phi_{\frac 12\lambda_j}-\phi_{\frac12
\lambda_j}\phi_{-\frac
12\lambda_i}=(-)^{n_j}\delta_{\lambda_i,\lambda_j},
 \]
 we see that, for partitions $\lambda$ and $\mu$,
we have the ortho-normality condition
\begin{gather}\label{orto}
\l\lambda \,|\,\mu \r = \delta_{\mu,\lambda}  ,\qquad \lambda, \mu \in
\rm OP.
\end{gather}

Let us introduce the following functions
 \begin{gather}\label{C-lambda}
\textsf{C}_\lambda(\bt):= \sigma\left(|\lambda\r  \right):= \l
0|\G(\bt)|\lambda\r ,\qquad \lambda\in \rm OP.
 \end{gather}
 These functions are weighted polynomials of weight $\frac 12|\lambda|$ in
 the variables
 $\bt:=(t_1,t_{\frac 12};t_3,t_{\frac 32};\dots)$  where $\deg t_j
 :=j $. For example
  \[
\C{(1)}=t_{\frac 12} ,\qquad \C{(1^2)}=\frac{t_1}{\sqrt{2}} .
  \]

 We evaluate $\textsf{C}_\lambda$ in case all (Grassmannian) odd variables vanish:
 $t_{n+\frac 12}=0$, $n=0,1,2,\dots$. (Recall that in this case we denote $\bt$ as
 $t=(t_1,t_3,\dots$).)

\begin{remark}\label{polynomials} Weighted polynomial functions are often
presented as symmetric functions of some variables
$x_1,\dots,x_N$, where the number of variables may be irrelevant.
In our case we put
 \begin{gather}\label{miwa}
t_{n} = \frac 1{n} \sum_{i=1}^N
\left(x_i^{n}-(-x_i)^{n}\right) ,\qquad n=1,2,3,\dots,
 \end{gather}
 where all even-labeled $t_n$ vanish. For $n$ odd we write
\[
t_{n} = \frac 2{n} \sum_{i=1}^N  x_i^{n} ,\qquad n=1,3,\dots,
\]
 where $x_i=\frac 1{z_i}$, $i=1,\dots,N$. Below by polynomial
 functions in Miwa variables we mean polynomial functions in the
 variables $x_i=\frac 1{z_i}$.
\end{remark}

  Then $\textsf{C}_\lambda(t)$ vanishes if $\ell(\lambda)$ is
  odd\footnote{We need to make
 dif\/ference between odd numbers and odd numbers in the Grassmannian sense. In the
last case we shall necessarily say Grassmannian odd number.}. If
$\ell(\lambda)$ is even, then by Wick's theorem
\[
\textsf{C}_\lambda(t)=\Hf\big[
\textsf{C}_{(\lambda_i,\lambda_j)}(t)\big] ,\qquad \lambda\in
\rm OP_e ,
\]
 where $\rm OP_e$ is the set of all partitions with even number of
 odd parts, and as we shall see
  \begin{gather}\label{b=s/c}
\textsf{C}_{(\lambda_i,\lambda_j)}(t)= \frac
1{d_{(\lambda_i|\lambda_j)}} s_{(n_i|n_j)}(t),
  \end{gather}
where $s_\lambda$ is the Schur function, and $(n|m)$ is a one-hook
partition in the Frobenius notations, see \cite{Mac}. Indeed,
f\/irst, from
 \[
[J_{2n-1}, \phi_{i}]=\phi_{i-2n+1}
 \]
 (cf.~\eqref{S6})
 it follows
 \[
\phi_{i}(t)
:=\G(t)\phi_{i}\G(t)^{-1}=\sum_{n=0}^\infty h_{n}(t)\phi_{i-n},
 \]
where $h_n$ are complete symmetric functions \cite{Mac}. Now from
(\ref{S1}), (\ref{S2}) we obtain
 \[
\l 0|\phi_{n_1+\frac 12}(t)\phi_{n_2+\frac 12}(t)|0\r
=\sum_{n=0}^{\lambda_1} (-1)^{n_2-n}h_{n}(t)h_{n_1+n_2+1-n}(t),
 \]
 while  Schur function evaluated on a one-hook partition is (see Chapter~I, \S~3, Example~9 in~\cite{Mac})
 \[
s_{(n_1|n_2)}({t})= h_{n_1+1}({t}) e_{n_2}({t})- h_{n_1+2}({t})
e_{n_2-1}({t})+\cdots+ (-)^{n_2}h_{n_1+n_2+1}({t})  ,
 \]
where $e_n$ are elementary symmetric functions. Then taking into
account that for $t=(t_1,t_3,\dots)$ of form~(\ref{miwa}) we get
the equality $h_m({t})\equiv e_m({t})$,  we obtain~(\ref{b=s/c}).

Thus we get
 \begin{gather}\label{C-Schur}
 \textsf{C}_\lambda(t)=
\begin{cases}
 \dfrac{1}{d_\lambda} \Hf\big[
s_{(n_i|n_j)}(t)\big] &\mbox{if}\ \
\ell(\lambda)\ \ \mbox{even},\\
0 &\mbox{if}\ \ \ell(\lambda)\ \ \mbox{odd}.
 \end{cases}
 \end{gather}
It follows from (\ref{C-Schur}) and from
$s_\lambda(-t)=(-1)^{|\lambda|}s_{\lambda^{tr}}(t)$ that
 \begin{gather}\label{C(-t)}
\C\lambda(-t)=(-1)^{\frac
12(|\lambda|+\ell(\lambda))}\C\lambda(t).
 \end{gather}

Next, if all Grassmannian odd variables  except $t_{\frac 12}$
vanish, we obtain
\[
 \textsf{C}_\lambda(t)=
 \begin{cases}
\dfrac{1}{d_\lambda} \Hf\big[
s_{(n_i|n_j)}(t)\big] &\mbox{if}\ \ \ell(\lambda)\ \ \mbox{even},\vspace{1mm}\\
t_{\frac 12} \dfrac{1}{d_\lambda} \Hf\big[ {\tilde S}\big]
&\mbox{if}\ \ \ell(\lambda)=2n-1 \ \ \mbox{odd},
 \end{cases}
\]
where ${\tilde S}$ is $2n \times 2n$ symmetric matrix
\[
{\tilde S}_{ij}={\tilde S}_{ji}:=
\begin{cases}
s_{(n_i|n_j)}(t) &\mbox{if}\ \ 1\le i<j \le 2n-1 , \\
s_{(n_i)}(t) &\mbox{if}\ \ 1\le i < j=2n .
 \end{cases}
\]
Recall that $t=(t_1,t_3,t_5,\dots)$ and that $\lambda_i$
are related to~$n_i$ via~(\ref{n-lambda}).

In general case, where the odd Grassmannian variables do not vanish,
we can see that $\C\lambda$ is of even Grassmannian parity in case
$\ell(\lambda)$ is even, and it is of odd Grassmannian parity in
case~$\ell(\lambda)$ is odd. We can write
\[
 \textsf{C}_\lambda(t)=
 \begin{cases}
\dfrac{1}{d_\lambda} \Hf\big[
s_{(n_i|n_j)}(t)\big]+C_{e} &\mbox{if}\ \ \ell(\lambda)\ \ \mbox{even},\vspace{1mm}\\
t_{\frac 12} \dfrac{1}{d_\lambda} \Hf\big[ {\tilde
S}\big]+C_{o} &\mbox{if}\ \ \ell(\lambda)=2n-1 \ \
\mbox{odd},
 \end{cases}
\]
 where $C_{e}$ and $C_{o}$ are polynomials in odd Grassmannian
 variables of the order~$\ell(\lambda)$, $C_{e}$ starts with
 quadratic terms, while $C_{o}$ starts with cubic ones. This
 follows from the consideration of~(\ref{prodphi}) in Subsection~\ref{Bose-Fermi}.

\subsection{Orthogonality}

 One can verify the equality, using (\ref{Gamma-bt})
 \begin{gather}\label{C-right-left}
\l 0|\,{\G}({\bt})|\lambda \r  =  \l \lambda|\,{\bar
\G}(-{\bt})| 0 \r,
 \end{gather}
where
 \begin{gather}\label{barGamma-bt}
{\bar\Gamma}({ \bt}):=e^{{\bar J}({\bf t})}  ,\qquad {\bar
J}({\bf t}) =  \sum_{0< i\in 1+2\Z}t_i J_{-i}+\sum_{0<
i\INZ12}   t_i J_{-i}
  \end{gather}
(keep the order in the products of Grassmannian odd variables,
which we label by semi-integer subscripts).

Thanks to (\ref{orto}) we can write
 \begin{gather}\label{left-coherent-state}
\l 0|\,{\G}({\bt}) = \l 0| + \sum_{\lambda\in \mathrm{OP}}
\textsf{C}_\lambda({\bt}) \l \lambda |.
 \end{gather}
On the other hand due to (\ref{C-right-left})
\[
{\bar\G}({\bt})\,|0\r = |0\r + \sum_{\lambda\in
\mathrm{OP}}\,|\lambda\r \textsf{C}_\lambda(-{\bt})
\]
and to (\ref{supercurrent}) we obtain
\[
\l 0|\,{\G}({\bt}){\bar\G}({\bar\bt})\,|0\r = e^{-\frac
12\sum\limits_{n=1,3,5,\dots} nt_n{\bar t}_n  -\frac 12
\sum\limits_{m=0,1,2,\dots } (-1)^{m}(m+\frac 12)t_{m+\frac 12}{\bar
t}_{m+\frac 12} },
\]
 where $\bt$ and ${\bar\bt}$ are two independent sets of
 variables:
  \[
{\bf t}=\left(t_1,t_{\frac 12};t_3,t_{\frac 32};t_5,t_{\frac
52};\dots \right) , \qquad {\bar\bt}=\left({\bar t}_1,{\bar
t}_{\frac 12};{\bar t}_3,{\bar t}_{\frac 32};{\bar t}_5,{\bar
t}_{\frac 52};\dots \right).
  \]
  On the other hand, due to (\ref{Gamma-bt}), (\ref{barGamma-bt})
and (\ref{orto})  we obtain the following analogue of Cauchy--Littlewood identity:
 \begin{gather}\label{CL-t}
e^{-\frac 12\sum\limits_{n=1,3,5,\dots}nt_n{\bar t}_n  -\frac 12
\sum\limits_{m=1,2,3,\dots }  (-1)^{m}(m+\frac 12)t_{m+\frac 12}{\bar
t}_{m+\frac 12} } = \sum_{\lambda\in
\mathrm{OP}} \textsf{C}_\lambda({\bt})  \textsf{C}_\lambda(-{\bar\bt}).
 \end{gather}

 From the last equality we obtain
\[
\G(\bt)=1+ \sum_{\lambda\in\mathrm{OP}} \textsf{C}_\lambda({\bt})
\textsf{C}_\lambda({\bf J}) ,\qquad
{\bar\G}(\bt)=1+ \sum_{\lambda\in\mathrm{OP}}  \C\lambda({\bar{\bf
J}})\textsf{C}_\lambda(-{\bar\bt}),
\]
 where
\begin{gather*}
 {\bf J}:=\left(2J_1,4J_{\frac 12};\frac 23 J_3,-\frac 43 J_{\frac 32};
 \frac 25 J_5,\frac 45 J_{\frac
 52};\frac 27 J_7,-\frac 47 J_{\frac 72};\dots\right) ,
\\
 {\bar{\bf J}}:=\left(-2J_{-1},-4 J_{-\frac 12};-\frac 23 J_{-3},
 \frac 43 J_{-\frac 32};-\frac 25 J_{-5},-\frac 45 J_{-\frac
 52};-\frac 27 J_{-7},
 \frac 47 J_{-\frac 72};\dots\right)
\end{gather*}
 and therefore
 \begin{gather}\label{b-state}
|\lambda\r =\textsf{C}_\lambda({\bar{\bf J}})\,|0\r ,\qquad \l
\lambda |=\l 0|\,\textsf{C}_\lambda({\bf J}).
 \end{gather}

Let $f(\bt)$ and $g(\bt)$ be series in the variables $\{t_i\}$. We
introduce the following scalar product
 \begin{gather}
\label{scalar-product}
 \l f , g\r : = \l 0|\,f({\bf J})
g({\bar{\bf J}})\,|0\r.
  \end{gather}
In particular due to (\ref{supercurrent})
\[
\l t_i, t_j\r = \l 0| J_{i}J_{-j} |0\r = -\frac j2
(-1)^{[-j-\frac 12]}\delta_{i,j},
\]
where $[a]$ denotes the integer part of $a$ (i.e.,
  $a=[a]+\epsilon$ where $0\le \epsilon \le 1$ for~$a<0$).

It follows from (\ref{b-state}) and (\ref{orto}) that
 the polynomials form an orthonormal basis in the scalar product
 (\ref{scalar-product}):
\[
\l  \textsf{C}_\lambda , \textsf{C}_\mu \r  = \l
0|\,\textsf{C}_\lambda({\bf J}) \textsf{C}_\mu({\bar{\bf
J}})\,|0\r = \delta_{\lambda,\mu}.
\]

\subsection[Polynomials  $\C\lambda$ in super Miwa variables]{Polynomials  $\boldsymbol{\C\lambda}$ in super Miwa variables}

If we want
to rewrite polynomials as symmetric functions symmetric with respect to the
action of symmetric group on pairs $(z_i,\zeta_i)$ of super Miwa
variables ${\bf z}=(z_1,\zeta_1,\dots ,z_k,\zeta_k)$
(\ref{s-Miwa}), we present $\C\lambda$ as
\[
\C\lambda({\bf z}) := \C\lambda(-[{\bf z}]) = \l
0|\,{:}e^{\Theta(z_1,\zeta_1)}{:}\, \cdots
\,{:}e^{\Theta(z_k,\zeta_k)}{:}\,|\lambda\r  \texttt{D}({\bf z})^{-1}
\]
in terms of super vertex operators (\ref{super-vertex-viaVphi}). Here
 \[
\texttt{D}({\bf z}) = \texttt{D}_k({\bf z}) := \l
0|\,{:}e^{\Theta(z_1,\zeta_1)}{:}\, \cdots
\,{:}e^{\Theta(z_k,\zeta_k)}{:}\,|0\r
 \]
is antisymmetric function with respect to the permutation
 of pairs $(z_i,\zeta_i)$, ${\texttt{D}}({\bf z})^{-1}$, see (\ref{D^-1}).
 Let us note that if $\ell(\lambda)>k$ the polynomial $C_\lambda$ does not
depend on the odd Grassmannian variables $\zeta_i$, and vanish for
$\ell(\lambda)$ odd.
Example. For $k=1$ we obtain
 \begin{gather}\label{C-z-zeta-k=1}
\C\lambda(z,\zeta)=
\begin{cases}
\dfrac 2{d_\lambda} z^{-\frac12|\lambda|}(2\ell(\lambda)-1)!!
&\mbox{if}\quad \ell(\lambda)\quad \mbox{even}, \vspace{1mm}\\
\dfrac{\zeta}{d_\lambda} z^{-\frac 12|\lambda|-\frac 12} &\mbox{if}\quad \ell(\lambda)=1,  \\
0&\mbox{if}\quad \ell(\lambda)>1,\, \mbox{odd},
 \end{cases}
 \end{gather}
 where we use the formula $s_{(n|m)}(x)=2x^{n+m+1}$.

In Miwa variables the Cauchy--Littlewood identity (\ref{CL-t}) is written as
\[
\sum_{\lambda\in
\mathrm{OP}}\,\textsf{C}_\lambda({\bf z})\, \textsf{C}_\lambda(-{\bar{\bf
z}})=
\prod_{i,j}\,\frac{1-z_i{\bar z}_j}{1+z_i{\bar
z}_j}
\left(1-
\zeta_i{\bar{\zeta}_j} \frac{1-z_i{\bar z}_j}{\left(1+z_i{\bar z}_j\right)^2} \right).
\]
\begin{proof} Using \eqref{s-Miwa}, namely
 \[
t_{2n+1}=\frac{2}{(2n+1)}\sum_{i=1}^k\frac 1{z_i^{2n+1}} ,\qquad
t_{m+\frac12} = 2\sum_{i=1}^k\frac{\zeta_i}{z_i^{m}}
  \]
we obtain
\begin{gather*}
e^{-\frac 12
\sum\limits_{m=1,2,3,\dots }  (-1)^{m}(m+\frac 12)t_{m+\frac 12}{\bar
t}_{m+\frac 12} } =
\prod_{i,j}
\left(1-2
\zeta_i{\bar{\zeta}_j}\sum\limits_{m=1,2,3,\dots }  (-1)^{m}\left(m+\frac 12\right)z_i^{-m}{\bar
z}_j^{-m} \right)
 \\
\qquad{}=\prod_{i,j}
\left(1+2
\zeta_i{\bar{\zeta}_j}D_{z_i}\sum\limits_{m=1,2,3,\dots }  (-1)^{m}z_i^{-m}{\bar
z}_j^{-m} \right)\\
\qquad{} =
\prod_{i,j}
\left(1+2
\zeta_i{\bar{\zeta}_j}D_{z_i}\left( -\frac{1}{z_i{\bar{z}_j}}\frac{1}{1+z_i^{-1}{\bar
z}_j^{-1}} \right) \right)
\\
\qquad{} =
\prod_{i,j}
\left(1-2
\zeta_i{\bar{\zeta}_j}D_{z_i}\left( \frac{1}{1+z_i{\bar
z}_j} \right) \right)=
\prod_{i,j}
\left(1-
\zeta_i{\bar{\zeta}_j} \frac{1-z_i{\bar z}_j}{\left(1+z_i{\bar z}_j\right)^2} \right),
\end{gather*}
where $D_z=\frac12(z\partial_z+\partial_z z)$ is the Euler operator.
\end{proof}

\subsection[Combinatorial meaning of $\textsf{C}_{\lambda}(t)=\textsf{C}_{\lambda}(1,0,0,\dots)$]{Combinatorial meaning of $\boldsymbol{\textsf{C}_{\lambda}(t)=\textsf{C}_{\lambda}(1,0,0,\dots)}$}

 Each partition with odd
parts may be presented as $\lambda=(1^{m_1}3^{m_3}5^{m_5}\cdots)$
where $m_i$ is the multiplicity of the number $i$ (that means that
the partition $\lambda$ contains the part equal to $i$  $m_i$
times). The length $\ell(\lambda)$ of the partition $\lambda$ is
equal to $\sum\limits_{i=1,3,5,\dots}m_i$, the weight is
$|\lambda|=\sum\limits_{i=1,3,5,\dots} i m_i$.

Let us visualize this, in a similar way as in the papers \cite{HO,LO}, as the one-dimensional semi-inf\/inite lattice
of cites (baskets) in our case numbered by odd positive integers.
A basket number~$i$ ($i=1,3,5,\dots$) contains $m_i$ identical
balls (and therefore the multiplicity~$m_i$ may be also called the
{\em occupation number}). Nonequivalent distributions of balls is
in one-to-one correspondence with partitions from the set OP.
(The length of a partition is equal to the number of balls, the
ratio of the weight of the partition and the length of the
partition may be considered as the location of the mass center of
the balls).

Let us consider the following discrete time random process
describing the creation of  $\lambda\in\mathrm{OP}$ or, the same,  of ball
conf\/igurations.  It starts with a given partition, say, $\mu$.
(The case where $\mu=0$ describes the conf\/iguration where all
baskets are empty  at time ${\textsc t}=0$). At each discrete time
instant ${\textsc t}=1,2,3,\dots$ one of the following  two possible events
occurs with equal probability (A)~either two balls are created in
the leftmost basket (basket number 1), or (B)~a~ball chosen at
random in any of baskets, say, in basket number $i$, is moved to
the nearest basket to the right (to the basket number $i+2$). It
is clear that at each time  step the weight of the related
partition increases: $|\lambda|\to |\lambda|+2$, thus
$|\lambda|=2{\textsc t}$.
 A problem is to f\/ind a number of ways to create a given
distribution $\lambda$ of the balls in baskets along the process
described above in ${\textsc t}=\frac 12|\lambda|$ steps. We
denote this number  $ \textsf{N}_{\mu\to\lambda}$.

Then we state that
\[
\C\lambda(1,0,0,0,\dots)=\left(\frac 12 \right)^{\frac
12\ell(\lambda)} \frac{1}{\left(\frac12 |\lambda|\right)!}
\frac{1}{d_\lambda} \textsf{N}_{0\to\lambda}
\]

The proof follows from
 \[
e^{J_{-1}}|0\r =|0\r+ \sum_{\lambda\in\rm OP_e}\,|\lambda
\r\C\lambda(-1,0,0,0,\dots)=\sum_{{\textsc
t}=0,1,2,\dots}\,\frac{1}{{\textsc t}!}\left(
J_{-1}\right)^{\textsc t}|0\r
 \]
and from the detailed consideration of the action of
  \[
J_{-1}=-\frac 12 \phi_{\frac 12}^2 +\phi_{-\frac 12}\phi_{\frac
32}-\phi_{-\frac 32}\phi_{\frac 52}+\cdots
 \]
 on
basis Fock vectors (\ref{basis-vectors}) (see Appendix~\ref{Realization}), and from (\ref{C(-t)}).

Now we write down the following formula for
$\C\lambda(1,0,0,\dots)$ obtained from (\ref{C-Schur}):
 \[
\C\lambda(1,0,0,\dots)= \frac{1}{d_\lambda} \prod_{i}\frac
1{n_i!n_j!}\Hf\left[ \frac{1}{n_i+n_j+1} \right],
 \]
where $n_i$ and $\lambda_i$ are related by (\ref{n-lambda}), which
yields
\[
\textsf{N}_{0\to\lambda}= 2^{\frac 12
\ell(\lambda)} \left(\frac 12
|\lambda|\right)! \left(\prod_{i=1,3,5,\dots} \frac{1}{m_i!}\right) \Hf\left[
\frac{1}{n_i!n_j!(n_i+n_j+1)} \right],
\]
 where we recall that $m_i$ and $n_i$ we def\/ined above:
 $\lambda=(1^{m_1}3^{m_3}5^{m_5}\cdots)$ and
 $\lambda=(\lambda_1,\dots,\lambda_{2k})$ where $\ell(\lambda)=2k$ is the length of
 the partition,
 $\lambda_i=2n_i+1$.

 For instance, take $\lambda=(1^{2k})$. Obviously there is only one
 way to create this conf\/iguration, $\textsf{N}_{0\to (1^{2k})}\,=1$. Indeed
 $\ell(\lambda)=|\lambda|=2k$, $n_i=0$,
 $i=1,\dots,2k$, and we obtain well-known identity
  \[
1 = \textsf{N}_{0\to(1^{2k})} = 2^k k! \frac{1}{(2k)!} (2k-1)!!.
  \]

At last let us mention that $\C\lambda(\bt)$ where $t_i=0$ except
$i=\frac 12,\, 1$ may be related to the numbers $\textsf{N}_{(1)\to
\lambda}$, where $\lambda\in\rm OP_o$.

\subsection[Polynomials $\textsf{C}_{\lambda/\mu}$]{Polynomials $\boldsymbol{\textsf{C}_{\lambda/\mu}}$}

We def\/ine skew polynomials $\textsf{C}_{\lambda/\mu}(\bt)$ as
follows (cf.~(\ref{Gamma-bt}) and~(\ref{barGamma-bt})):
\[
\textsf{C}_{\lambda/\mu}(\bt) := \l \mu|\G(\bt)|\lambda \r = \l
\lambda|{\bar\G}(-\bt)|\mu \r.
\]
 From this def\/inition $\textsf{C}_{\lambda/\mu}$ vanishes
unless $\mu \subseteq \lambda$. The same may be written as
 \begin{gather*}
\G(\bt) = \sum_{\lambda,\mu} \C{\lambda/\mu}(\bt)\,|\mu \r \l
\lambda|, \qquad
{\bar\G}(\bt) = \sum_{\lambda,\mu} \C{\lambda/\mu}(-\bt)\,|\lambda
\r \l \mu|.
\end{gather*}

 If $\bt=\bt'+\bt''$ then $\G(\bt)=\G(\bt')\G(\bt'')$ by inserting the unity
 operator $\sum\limits_{\lambda\in P}|\lambda\r \l \lambda|$ between $\G(\bt')$ and $\G(\bt'')$
 we obtain
\[
\textsf{C}_{\lambda}(\bt) = \sum_{\mu}
 \textsf{C}_{\lambda/\mu}(\bt')\textsf{C}_{\mu}(\bt'').
\]
  This property is quite similar to the property of the Schur
  functions (see~(5.9) in~I of~\cite{Mac}).

  One may relate $\C{\lambda/\mu}$ to the numbers $
  \textsf{N}_{\mu\to\lambda}$ described in the previous
  subsection.

\subsection[CKP tau function and polynomials $\textsf{C}_\lambda$]{CKP tau function and polynomials $\boldsymbol{\textsf{C}_\lambda}$}\label{CKPpolynomials}

First of all we note that $\textsf{C}_\lambda(\bt)$ is not a solution of the Hirota bilinear equations, and, therefore
is not a CKP tau function.
However, due to (\ref{left-coherent-state}) CKP tau functions~(\ref{superTau}) are series in $\textsf{C}_\lambda(\bt)$ as
follows
\[
\tau(\bt)=\sum_{\lambda\in\rm OP} g_\lambda \textsf{C}_\lambda(\bt),
\]
 where
 \[
g_\lambda = \l \lambda |\,g\,|0\r.
 \]

%{\bf Example 1.}
\begin{example}
Take $g=\exp \sum\limits_{i>0} e^{-2U_i}\phi_i^2$.
Then
 \begin{gather}\label{ex1}
\tau(\bt) = \sum_{\lambda} e^{-U_\lambda}
\C\lambda(\bt)d_\lambda \prod_{i=1,3,5,\dots} \frac{1}{k_i !},
 \end{gather}
where the sum ranges over all $\lambda\in\rm OP$ whose parts have even
multiplicities, $m_i=:2k_i$, i.e., of the form $\lambda
=\left(1^{2k_1}3^{2k_3}5^{2k_5}\cdots \right)$.
  The numbers $U_\lambda$ are def\/ined as
 \begin{gather}\label{U-lambda}
 U_\lambda := \sum_{i=1}^k U_{\frac 12 \lambda_i}.
 \end{gather}

The right-hand side of~(\ref{ex1}) may be compared to  sums over
partitions in~\cite{HLO} and in~\cite{sQQ} dealing with tau
functions of neutral and charged BKP hierarchies, respectively.
\end{example}

%{\bf Example 2.}
\begin{example}
Given a symmetric matrix $A$ and a partition
$\lambda$ introduce numbers $A_\lambda$ according to the formula
\[
e^{\sum\limits_{n,m > 0}    x_nA_{nm}x_m} = \sum_{\lambda\in
P_{\rm ev}} A_\lambda x_\lambda,
\]
  where $\lambda=(\lambda_1,\dots,\lambda_k)$ is a partition,
  and $x_\lambda=x_{\lambda_1}\cdots x_{\lambda_k}$, and where
  $P_{\rm ev}$ is a set of all partitions with even number of parts.

  Then taking
\[
e^{\sum\limits_{n,m > 0}    \phi_{\frac n2}A_{nm}\phi_{\frac m2}},
\]
 which is an exponential of a quadratic form of  creation operators, we obtain
\[
\tau(\bt) = \l 0|\,\Gamma(\bt) g\,|0\r  = \sum_{\lambda\in
P_{\rm ev}} A_\lambda \C{\lambda}(\bt).
\]

Next we may write the CKP tau function as a double series over
partitions
\[
\tau(\bt,{\bar t})=\sum_{\lambda,\mu\in
\rm OP} g_{\lambda,\mu} \textsf{C}_\lambda(\bt)\C\mu({\bar t}),
\]
 where
 \[
g_{\lambda,\mu} = \l \lambda |\,g\,|\mu\r.
 \]
\end{example}

%{\bf Example 3.}
\begin{example}
\[
\tau(\bt,U,{\bar t})=\l 0|\,\G(\bt)\mathbb{T}(U){\bar \G}({\bar
t})\,|0\r=1+\sum_{\lambda\in
\rm OP} e^{-U_\lambda}\textsf{C}_\lambda(\bt)\textsf{C}_\lambda(-{\bar
t}),
\]
  where $U=\big(U_\frac 12,U_\frac 32,U_\frac 52,\dots\big)$ is a set of
  constants, and
  \[
\mathbb{T}(U)   := \exp \sum_{0< i\in \frac 12 +\mathbb{Z}}
(-1)^{i+\frac 12}U_i\varphi_i\varphi_{-i},
  \]
  where the numbers $U_\lambda$ are def\/ined by (\ref{U-lambda}).

This example may be compared with the results of \cite{pd22} and~\cite{q2} devoted to TL and neutral BKP hypergeometric tau
functions.

We can specify ${\bar t}$ in such a way that ${\bar t}_{i}=\frac
2i z^i$, $i=1,3,5, \dots,$ and use~(\ref{C-z-zeta-k=1}). We thus obtain
\[
\tau(\bt,U,{\bar t}(z))=1+\sum_{\lambda\in {\rm OP}_{\rm ev}} f_\lambda
e^{-U_\lambda}\textsf{C}_\lambda(\bt)  ,\qquad
f_\lambda:=(-1)^{\frac 12(|\lambda|+\ell(\lambda))}\frac
2{d_\lambda} z^{\frac12|\lambda|} (2\ell(\lambda)-1)!!.
\]
\end{example}

\appendix

\section{Examples of CKP tau functions}\label{Examples}

(I) One-soliton tau function is
\[
\l 0|\, \Gamma(t) e^{\frac
a2\phi(p)\phi(q)}\,|0\r = \left(1-\frac{a}{p+q}e^{\sum_{}\left(p^n+q^n
\right)t_n} \right)^{-\frac 12}.
\]

(II) Another example is as follows
\[
\l 0|\, \Gamma(t) e^{ a\phi_{\frac 12}^2}\,|0\r = \left(1-at_1
\right)^{-\frac 12},
\]
 which may be also viewed as a solution of a heat equation
 \[
\frac{\partial f(z_{\frac 12},t_1)}{\partial
{t_1}}=\frac{\partial^2 f(z_{\frac 12},t_1)}{\partial z_{\frac
12}^2} ,\qquad f(z_{\frac 12},0)= \exp  a z_{\frac 12}^2,
 \]
which is
  \[
f(z_{\frac 12},t_1)=\exp t_1\partial^2_{\frac 12}\cdot \exp a
z^2_{\frac 12} = \frac{1}{\sqrt{1-at_1}} \exp \frac{ a z_{\frac
12}^2 } {1-at_1}
  \]
in the origin.

\section[A realization of the algebra of free bosons (\ref{S1})-(\ref{S2BASIS})]{A realization of the algebra of free bosons (\ref{S1})--(\ref{S2BASIS})}\label{Realization}

 The simplest way to understand the action of $J_{-1}$ on a
basis Fock vector may be as follows. Consider the realization of
free boson algebra (\ref{S1}) via dif\/ferentiation operators in
auxiliary variables $z_i$:
\[
\phi_{m+\frac12}=z_{m+\frac12},\qquad
\phi_{-m-\frac12}=(-)^{m}\partial_{m+\frac12} ,\qquad m\ge 0,
\]
 while the Fock space $F$ may be viewed as polynomial functions in
 the auxiliary variables $z_i$
\[
|(1^{m_1}3^{m_3}5^{m_5}\cdots)\r \,d_\lambda = z_{\frac
12}^{m_1}z_{\frac 32}^{m_3}z_{\frac 52}^{m_5}\cdots.
\]

The action of $J_{-1}$ on a basis Fock vector yields the following
 \begin{gather*}
\left( -\frac 12z_{\frac12}^2 + z_{\frac32} \partial_{\frac12}+
z_{\frac52}
\partial_{\frac32}+\cdots  \right)
z_{\frac 12}^{m_1}z_{\frac 32}^{m_3}z_{\frac 52}^{m_5}\cdots
\\
\qquad{}=-\frac 12 z_{\frac 12}^{m_1+2}z_{\frac 32}^{m_3}z_{\frac
52}^{m_5}\cdots + m_1 z_{\frac 12}^{m_1-1}z_{\frac
32}^{m_3+1}z_{\frac 52}^{m_5}\cdots   +  m_3 z_{\frac
12}^{m_1}z_{\frac 32}^{m_3-1}z_{\frac 52}^{m_5+1}\cdots  +
\cdots.
 \end{gather*}
As we see the action of $J_{-1}$ on a basis vector results in
linear combination of basis vectors. The f\/irst term in the right-hand
side of the last equality may be related to the event (A)
(creation of a pair of balls in the basket~1. This event is
accomplished by the multiplication by the factor~$-\frac 12$. Thus
we get the general prefactor $(-1)^{\ell(\lambda)}$ for all
conf\/igurations with $\ell(\lambda)$ number of the balls. Other
terms in the right-hand side may be related to the event (B) where
a chosen ball is moved to the right neighbor basket, each
possibility has the weight~1. The factors $m_i$ describe the fact
that we can chose any of~$m_i$ balls in the basket~$i$ to move
them to the right.

\section{Bosonic KP tau function}\label{BosonicKP}

Here we shall brief\/ly describe a bosonic KP tau function and its
symplectic reduction to the bosonic CKP tau function.

Consider the following bosonic operators
\[
[p_i,\,q_j] = \delta_{i,j} ,\qquad i, j\in \frac 12+\mathbb{Z} .
\]
The right and left Fock spaces will be def\/ined via
\[
p_i|0\r=q_{-i}|0\r = 0 = \l 0|q_i=\l 0|p_{-i} ,\qquad i<0 .
\]

Let us def\/ine the expression $p_iq_j-\l 0|p_iq_j |0\r$  by $E_{ij}={:}p_iq_j{:}$. These $E_{ij}$ may be considered
as generators of $gl(\infty)$ algebra with a negative level.

 Let us notice that for
  \begin{gather}\label{KPbosonic g}
g:=\exp \sum_{i,j\in \frac 12+\mathbb{Z}} A_{ij}\,{:}p_iq_j{:}
  \end{gather}
  we have
\[
\sum_{i\in\frac12 + \mathbb{Z}}gp_i\otimes
gq_i = \sum_{i\in\frac12 + \mathbb{Z}}p_ig\otimes q_ig .
\]

Bosonic KP tau function may be def\/ined as
 \[
  \tau^{KP}(\bt):=\l 0|\Gamma(\bt) g |0\r ,\qquad \Gamma(\bt):=\exp \sum_{n=1}^\infty t_n \sum_{k\in\mathbb{Z}} p_kq_{k+n} ,
 \]
where $\bt=(t_1,t_2,\dots)$ is a set of higher times. This tau function may be considered as a particular case
of a supersymmetric KP tau function \cite{Kac-Leur}.

If we ask $g$ of \eqref{KPbosonic g} to be invariant under $Sp(\infty)$ group then $g$ may be expressed in terms of
the so-called symplectic bosons.

{\bf Symplectic bosons.} Now between the bosons
\[
\phi_{i}=\frac1{\sqrt{2}}\left(p_{i}-(-1)^{i+\frac12}q_{-i-1}\right) ,\qquad
\hat{\phi}_{i}=\frac1{\sqrt{2}}\left(p_{i}+(-1)^{i+\frac12}q_{-i-1}\right)
\]
there are the following relations, cf.~\eqref{S1}, \eqref{S2}:
\[
 [\phi_{i},\phi_j]=[\hat{\phi}_{j},\hat{\phi}_i]=(-)^{j-\frac{1}{2}}\delta_{i,-j}
\qquad\hbox{and}\qquad
 [\phi_{i},\hat{\phi}_j]=0.
\]

It is known (see~\cite{Kac-Leur}) that quadratic expressions $Z_{ij}={:}\phi_i\phi_j{:}$ ordered via \eqref{normal ordering} may
be considered as a realization for the generators of the $c_\infty$ algebra with a negative level (this fact may be verif\/ied
with the help of~\eqref{S1} and~\eqref{normal ordering}). The same is true for $\hat{Z}_{ij}={:}\hat{\phi}_i\hat{\phi}_j{:}$.

It is straightforward to show that if $g$ of \eqref{KPbosonic g} is invariant under the  $Sp(\infty)$ group then $g=g_1\hat{g}_1$, where
$g_1=\exp \Big(\sum\limits_{i,j\in \frac 12+\mathbb{Z}} A_{ij}\,{:}\phi_i\phi_j{:}\Big)$ and
$\hat{g}_1=\exp\Big(\sum\limits_{i,j\in \frac 12+\mathbb{Z}} A_{ij}\,{:}\hat{\phi}_i\hat{\phi}_j{:} \Big)$; then, if all $t_{2n-1}=0$, the KP
bosonic tau function may be factorized: $\tau^{\rm KP}=\left(\tau^{\rm CKP} \right)^{-2}$, where $ \tau^{\rm CKP}$ is of form~\eqref{pr77},
 $\alpha=0$ (see also~(\ref{superTau})).

\subsection*{Acknowledgements}

The authors thank J.~Harnad for many discussions on related topics.
This work has been partially supported by
the European Union through the FP6 Marie Curie RTN
ENIGMA (Contract no.~MRTN-CT-2004-5652), the European Science
Foundation Program MISGAM, by RFBR grant 11-01-00440-a, and by
JSPS-RFBR grant 10-01-92104 JF.

\pdfbookmark[1]{References}{ref}
\LastPageEnding


\begin{thebibliography}{99}
\footnotesize\itemsep=0pt

\bibitem{AL}
Aratyn H., van~de Leur J.W., The {CKP} hierarchy and the {WDVV} prepotential, in
  Bilinear Integrable Systems: from Classical to Quantum, Continuous to
  Discrete, \href{http://dx.doi.org/10.1007/978-1-4020-3503-6_1}{\textit{NATO Sci. Ser.~II Math. Phys. Chem.}}, Vol.~201, Springer,
  Dordrecht, 2006, 1--11, \href{http://arxiv.org/abs/nlin.SI/0302004}{nlin.SI/0302004}.

\bibitem{DJKM-CKP}
Date E., Jimbo M., Kashiwara M., Miwa T., Transformation groups for soliton
  equations. {III}.~{O}perator approach to the {K}adomtsev--{P}etviashvili
  equation, \href{http://dx.doi.org/10.1143/JPSJ.50.3806}{\textit{J.~Phys. Soc. Japan}} \textbf{50} (1981), 3806--3812.

\bibitem{DJKMVI}
Date E., Jimbo M., Kashiwara M., Miwa T., Transformation groups for soliton
  equations. {VI}.~{KP} hierarchies of orthogonal and symplectic type,
  \href{http://dx.doi.org/10.1143/JPSJ.50.3813}{\textit{J.~Phys. Soc. Japan}} \textbf{50} (1981), 3813--3818.

\bibitem{DJKM}
Date E., Kashiwara M., Jimbo M., Miwa T., Transformation groups for soliton
  equations, in Nonlinear Integrable Systems~-- Classical Theory and Quantum
  Theory ({K}yoto, 1981), World Sci. Publishing, Singapore, 1983, 39--119.

\bibitem{HO}
Harnad J., Orlov A.Yu., Fermionic construction of tau functions and random
  processes, \href{http://dx.doi.org/10.1016/j.physd.2007.05.011}{\textit{Phys.~D}} \textbf{235} (2007), 168--206,
  \href{http://arxiv.org/abs/0704.1157}{arXiv:0704.1157}.

\bibitem{HLO}
Harnad J., van~de Leur J.W., Orlov A.Yu., Multiple sums and integrals as neutral BKP
  tau functions, \href{http://dx.doi.org/10.1007/s11232-011-0077-z}{\textit{Theoret. and Math. Phys.}} \textbf{168} (2011),
  951--962, \href{http://arxiv.org/abs/1101.4216}{arXiv:1101.4216}.

\bibitem{IKO}
Ishikawa M., Kawamuko H., Okada S., A Pfaf\/f\/ian--Hafnian analogue of Borchardt's
  identity, \href{http://arxiv.org/abs/math.CO/0408364}{math.CO/0408364}.

\bibitem{KacVAfB}
Kac V., Vertex algebras for beginners, \textit{University Lecture Series},
  Vol.~10, 2nd ed., American Mathematical Society, Providence, RI, 1998.

\bibitem{Kac-Leur}
Kac V.G., van~de Leur J.W., Super boson-fermion correspondence of type~$B$, in
  Inf\/inite-Dimensional {L}ie Algebras and Groups ({L}uminy-{M}arseille, 1988),
  \textit{Adv. Ser. Math. Phys.}, Vol.~7, World Sci. Publ., Teaneck, NJ, 1989,
  369--406.

\bibitem{KL}
Kac V.G., van~de Leur J.W., The {$n$}-component {KP} hierarchy and
  representation theory, \href{http://dx.doi.org/10.1063/1.1590055}{\textit{J.~Math. Phys.}} \textbf{44} (2003),
  3245--3293, \href{http://arxiv.org/abs/hep-th/9308137}{hep-th/9308137}.

\bibitem{Mac}
Macdonald I.G., Symmetric functions and {H}all polynomials, 2nd ed., Oxford
  Mathematical Monographs, The Clarendon Press Oxford University Press, New
  York, 1995.

\bibitem{Nimmo}
Nimmo J.J.C., Hall--{L}ittlewood symmetric functions and the {BKP} equation,
  \href{http://dx.doi.org/10.1088/0305-4470/23/5/018}{\textit{J.~Phys.~A: Math. Gen.}} \textbf{23} (1990), 751--760.

\bibitem{q2}
Orlov A.Yu., Hypergeometric functions related to Schur $Q$-polynomials and the
  BKP equation, \href{http://dx.doi.org/10.1023/A:1027370004436}{\textit{Theoret. and Math. Phys.}} \textbf{137} (2003),
  1574--1589, \href{http://arxiv.org/abs/math-ph/0302011}{math-ph/0302011}.

\bibitem{sQQ}
Orlov A.Yu., Shiota T., Takasaki K.,
Pfaffian structures and certain solutions to BKP hierarchies. I.~Sums over partitions,
\href{http://arxiv.org/abs/1201.4518}{arXiv:1201.4518}.



\bibitem{pd22}
Orlov A.Yu., Scherbin D.M., Multivariate hypergeometric functions as
  {$\tau$}-functions of {T}oda lattice and {K}adomtsev--{P}etviashvili
  equation, \href{http://dx.doi.org/10.1016/S0167-2789(01)00158-0}{\textit{Phys.~D}} \textbf{152/153} (2001), 51--65,  \href{http://arxiv.org/abs/math-ph/0003011}{math-ph/0003011}.

\bibitem{Sa}
Sato M., Soliton equations as dynamical systems on an inf\/inite dimensional
  Grassmann manifolds, in Random Systems and Dynamical Systems (Kyoto, 1981),
  \textit{RIMS Kokyuroku}, Vol.~439, Kyoto, 1981, 30--46.

\bibitem{Tinit}
Takasaki K., Initial value problem for the {T}oda lattice hierarchy, in Group
  Representations and Systems of Dif\/ferential Equations ({T}okyo, 1982),
  \textit{Adv. Stud. Pure Math.}, Vol.~4, North-Holland, Amsterdam, 1984,
  139--163.

\bibitem{LO}
van~de Leur J.W., Orlov A.Yu., Random turn walk on a half line with creation of
  particles at the origin, \href{http://dx.doi.org/10.1016/j.physleta.2009.02.068}{\textit{Phys. Lett.~A}} \textbf{373} (2009),
  2675--2681, \href{http://arxiv.org/abs/0801.0066}{arXiv:0801.0066}.

\bibitem{You}
You Y., Polynomial solutions of the {BKP} hierarchy and projective
  representations of symmetric groups, in Inf\/inite-Dimensional {L}ie Algebras
  and Groups ({L}uminy-{M}arseille, 1988), \textit{Adv. Ser. Math. Phys.},
  Vol.~7, World Sci. Publ., Teaneck, NJ, 1989, 449--464.

\end{thebibliography}
\end{document}